\documentclass[twocolumn, notitlepage, 10pt, aps, floatfix, showpacs, prb, citeautoscript, superscriptaddress]{revtex4-1}
\usepackage{graphicx}
\usepackage{amsmath, amssymb}
\usepackage{bm}
\usepackage{xcolor}
\usepackage[colorlinks, citecolor={blue!50!black}, urlcolor={blue!50!black}, linkcolor={red!50!black}]{hyperref}
\usepackage{bookmark}
\usepackage[version=4]{mhchem}
\usepackage{microtype}
\usepackage[load=physical,load=abbr, range-units=single]{siunitx}

\newcommand{\pmat}[1]{\begin{pmatrix}#1\end{pmatrix}}
\newcommand{\comment}[1]{}
\newcommand{\sgn}{\textrm{sgn}}

\begin{document}
\title{Dissipation-enabled fractional Josephson effect}

\author{Doru Sticlet}
\affiliation{Kavli Institute of Nanoscience, Delft University of Technology, P.~O.~Box 4056, 2600 GA Delft, The Netherlands}
\affiliation{National Institute for Research and Development of Isotopic and Molecular Technologies, 67-103 Donat, 400293 Cluj-Napoca, Romania
}
\author{Jay D. Sau}
\affiliation{Department of Physics, Condensed Matter theory center and the Joint Quantum Institute, University of Maryland, College Park, MD 20742}
\author{Anton Akhmerov}
\affiliation{Kavli Institute of Nanoscience, Delft University of Technology, P.~O.~Box 4056, 2600 GA Delft, The Netherlands}

\begin{abstract}
The anomalous $4\pi$-periodic ac Josephson effect, a hallmark of topological Josephson junctions, was experimentally observed in a quantum spin Hall insulator.
This finding is unexpected due to time-reversal symmetry preventing the backscattering of the helical edge states and therefore suppressing the $4\pi$-periodic component of the Josephson current.
Here, we analyze the two-particle inelastic scattering as a possible explanation for this experimental finding.
We show that a sufficiently strong inelastic scattering restores the $4\pi$-periodic component of the current beyond the short Josephson junction regime.
Its signature is an observable peak in the power spectrum of the junction at half the Josephson frequency.
We propose to use the exponential dependence of the peak width on the applied bias and the magnitude of the dc current as means of verifying that the inelastic scattering is indeed the mechanism responsible for the $4\pi$-periodic signal.
\end{abstract}

\maketitle

\section{Introduction}
\label{sec:intro}
\begin{figure}[t]
\includegraphics[width=0.81\columnwidth]{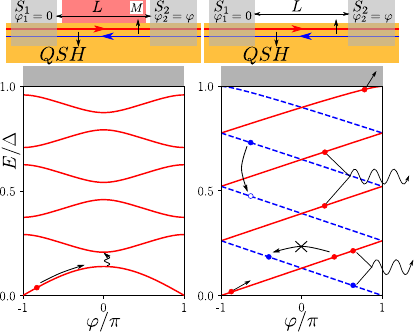}
\caption{(Color online) Josephson junctions created at the edge of a QSH insulator and the corresponding Andreev bound-state spectrum.
Two superconducting leads $S_{1,2}$ with a phase difference $\varphi$ connect the helical edge states of a QSH insulator.
(Left) Conventional setup for a topological junction. A magnetic material $M$ couples the counter-propagating edge states.
For example, a quasiparticle inhabits the lowest Andreev state.
At fixed parity, the ground state is $4\pi$ periodic, but Landau-Zener transitions (wavy line), which excite higher states, may destroy the $4\pi$-periodic effect.
(Right) The model studied in this paper, where two-particle dissipation generates a $4\pi$-periodic effect.
Time-reversal symmetry prohibits elastic scattering of single quasiparticles between counter-propagating edge states.
Several dissipative processes are allowed: (i) excitation of a particle at Fermi level and loss of quasiparticle into the continuum states, (ii) single-particle relaxation, and (iii) two-particle relaxation with pairwise annihilation of copropagating and antipropagating quasiparticles and emission of a photon.
}
\label{fig:sketch}
\end{figure}

\comment{QSH insulators have Majoranas and 4 pi Josephson effect.}
Quantum spin Hall (QSH) insulators~\cite{Kane2005, Bernevig2006, Koenig2007} are a promising platform for creation and manipulation of Majorana bound states.
The Majorana bound states arise in the topological edge states of QSH insulators, at the interface between the regions proximitized by a conventional $s$-wave superconductor and the regions with a magnetic gap~\cite{Fu2009}.
Since a pair of Majorana states in a Josephson junction gives rise to an anomalous $4\pi$-periodic Josephson effect,~\cite{Kitaev2001, Kwon2013} a magnetic Josephson junction in a QSH insulator should exhibit this phenomenon (see Fig.~\ref{fig:sketch}).
Recent experimental progress~\cite{Bocquillon2017, Deacon2017} has shown signatures of $4\pi$ periodicity in topological SNS (superconductor-normal metal-superconductor) junctions based on the QSH \ce{HgTe}/\ce{CdTe} quantum wells proximitized with \ce{Al} superconducting leads.

\comment{This effect was observed without magnetic insulators, and this is unexpected because the existing theory predicts either 2 or 8 pi in that case.}
Unexpectedly, the experimental observation of the anomalous Josephson effect did not require magnetic insulators, or any other source of time-reversal symmetry breaking.
This is unexpected since, as explained in Ref.~\onlinecite{Fu2009} and in later works, the time-reversal symmetry protects the finite-energy Andreev level crossings and results in a perfect pumping of quasiparticles to the energies above the superconducting gap, ultimately giving rise to a $2\pi$-periodic occupation of Andreev states and the conventional ac Josephson effect.
Extending this single-particle picture with elastic scattering due to interactions~\cite{Zhang2014, Orth2015} or to interaction with spinful impurities~\cite{Peng2016, Hui2017} removes the protection of the higher level crossings by allowing simultaneous elastic backscattering of two Andreev states.
Nevertheless, this leads to an $8\pi$-periodic, and not a $4\pi$-periodic Josephson effect.
Further phenomenological studies, where the Josephson junctions host both $2\pi$ and $4\pi$ currents, were done in the resistively shunted junction model.~\cite{Dominguez2017}

\comment{We show how the two-particle inelastic scattering restores the periodicity of Josephson effect.}
The inconsistency between the experimental observations and the theoretical predictions is the starting point of our investigation.
We propose and analyze the generation of a $4\pi$-periodic Josephson current due to the inelastic two-particle relaxation (a similar idea was mentioned in Ref.~\onlinecite{Peng2016}).
We show that if the dissipation is sufficiently strong and the Josephson junction contains several levels to enable the pairwise annihilation of the co-propagating quasiparticles (see Fig.~\ref{fig:sketch}), the fractional Josephson effect develops.
In the limit of large relaxation rate, the two-particle relaxation forces the Josephson junction to always stay in the lowest-energy state of a given fermion parity, and therefore results in a deterministic $4\pi$-periodic current-phase relationship.
Going beyond the limit of strong relaxation, we show that the fractional peak survives as long as the rate of losing quasiparticles into the continuum spectrum is much lower than the Josephson frequency.
In this regime, despite relaxation events taking place at arbitrary times, the correlation time of the fermion parity stays long, and guarantees the sharpness of the fractional peak.

\comment{While experimental detection of the inelastic scattering rates is hard, we predict two distinguishing features of this mechanism: the exponential peak width and the strongly nonlinear IV.}
The $4\pi$-periodic Josephson peak may appear also in a topologically trivial junction due to several reasons.~\cite{Sau2012, Houzet2013, Sau2017}
In order to distinguish the relaxation-enabled fractional Josephson effect from the one appearing due to alternative origins, we analyze the $I(V)$ characteristic of the Josephson junction as well as the shape of the fractional emission peak.
First, we find that there should be a critical Josephson frequency above which the fractional Josephson peak disappears.
This happens when the relaxation rate is not strong enough to ensure isolation of Andreev states from continuum states.
Because of the protected crossings in the spectrum, the inelastic processes become available already in the adiabatic limit, resulting in a linear (and square-root) voltage-dependent dc current already at low Josephson frequency, in contrast to the Landau-Zener tunneling processes that produce an exponentially vanishing dc current.
The low-frequency saturation of the amount of dissipated energy is a unique characteristic of this topological junction.
Finally, we predict that the width of the fractional peak should decrease exponentially with the Josephson frequency, and therefore with the applied bias voltage in the regime where the dc current is linearly or square-root varying with voltage.

The organization of the paper is the following.
In Sec.~\ref{sec:model} we present the model for the QSH Josephson junction.
The section also describes the rate-equation approach used to characterize the system dynamics and the basic tools used to extract the power spectrum of the junction.
Section~\ref{sec:1lev} treats the limit case of short junctions where two-particle relaxation takes place only at odd $\varphi/\pi$.
Section~\ref{sec:Nlev} extends the study to long junctions with many levels.
Here, we investigate two models for two-particle dissipation, one in which the relaxation rates are energy and time independent and one in which rates have a cubic dependence on excited quasiparticle energies.
In the latter, two-particle relaxation is facilitated by the junction coupling to an electromagnetic bath (see Appendix~\ref{app:circ}).
Finally, Sec.~\ref{sec:conc} holds the concluding remarks of the study.

\section{Model}
\label{sec:model}
\subsection{Spectrum of Andreev bound states}
The Josephson current in the QSH junction depends on the Andreev bound states in the junction and their occupation.
For this reason, we start by reviewing the Andreev bound-state spectra of such junctions.
Specifically, we consider ideal QSH edges connected by two superconducting leads placed at $\pm L/2$.
The setup is that of a symmetric SNS junction where the two leads have a relative superconducting phase difference $\varphi$.
The helical states of the QSH insulator carry a current between the leads over a distance $L$.
Therefore, the Thouless energy associated to the quasiparticle dwell time in the junction is $E_T=\hbar v/L$, with $v$ the Fermi velocity of the helical states.

The effective Hamiltonian for the Josephson junction at one edge of the QSH insulator reads as
\begin{equation}\label{ham}
H=(-i\hbar v\sigma_3\partial_x-\mu)\tau_3
+\Delta(x)e^{i\varphi(x)\tau_3}\tau_1,
\end{equation}
with $\bm\sigma$ and $\bm\tau$ the Pauli matrices in spin and, respectively, particle-hole space.
The Fermi velocity $v$ and chemical potential $\mu$ depend on material parameters.
The superconducting gap $\Delta$ is real, homogeneous, and present only in the superconducting leads $\Delta(x)=\Delta\Theta(L/2+x)\Theta(L/2-x)$, with $\Theta$ the Heaviside step function.
Since the physics depends only on the relative phase difference $\varphi$ between the superconducting leads, we  choose $\varphi(x)=\varphi\Theta(x-L/2)$.

The Andreev bound-state spectrum is determined by solving the Schr\"odinger equation with Hamiltonian~\eqref{ham} at fixed $\varphi$ using appropriate boundary conditions at the interface between the QSH insulator and the superconducting leads:
\begin{equation}\label{spect}
\arccos\big(\frac{\varepsilon^\pm_n}{\Delta}\big)\pm\frac\varphi 2-\frac{\varepsilon^\pm_n}{E_T}=n\pi,
\end{equation}
with $\pm$ standing for the right- (say spin up) and left-moving (spin down) eigenstates (see Fig.~\ref{fig:sketch}).
The above formula reproduces the short-junction spectrum by taking the limit $E_T\gg \Delta$:
\begin{equation}\label{sj}
\varepsilon^\pm=\mp(-1)^k\Delta\cos(\varphi/2),\quad\varphi
\in 2\pi[k,k+1),
\end{equation}
with $k$ an integer.
In the opposite (long-junction) limit $E_T\ll\Delta$, the spectrum is linearized:
\begin{equation}\label{eigs}
\frac{\varepsilon_n^\pm}{\pi E_T}=\bigg(
n+\frac1 2
\bigg)\pm \frac{\varphi}{2\pi}.
\end{equation}
Here, we neglect corrections to the current of the order $eE_T/\hbar$ in the low-dissipation/high-voltage regime, where Andreev levels with $E\approx \Delta$ become occupied.
In the long junction there are approximately $2N$ positive levels, $N=\lfloor \Delta/\pi E_T\rfloor$, which may be filled by quasiparticles (here and later $\lfloor x\rfloor$ is the floor function).

The electric current carried by an Andreev level is $ 2e\partial_\varphi\varepsilon^\pm_n/\hbar= \pm ev/L$.
The ground-state energy of the junction is obtained by summing over all negative levels $E_{gs}=\frac{1}{2}\sum_{\sigma=\pm,n}\varepsilon_{n}^{\sigma} \Theta(-\varepsilon_{n}^{\sigma})$~\cite{Bardeen1969, Beenakker1991, Chtchelkatchev2003}.
The supercurrent contribution from the ground state $I_{gs}=2e\hbar^{-1}\partial_\varphi E_{gs}$ follows readily, yielding a piecewise linear dependence of the current on the superconducting phase difference~\cite{Ishii1970}:
\begin{equation}\label{igs}
\frac{I_{gs}}{i_0}=\frac{\varphi}{2\pi}
-\Big\lfloor\frac{\varphi+\pi}{2\pi}\Big\rfloor,
\quad i_0=\frac{ev}{L}.
\end{equation}
The ground-state current is $2\pi$-periodic and odd in phase.
The sawtooth shape of the current exhibits jumps of height $i_0$ associated to the relaxation of a quasiparticle at odd $\varphi/\pi$, with $i_0$ the current carried by a single Andreev state.

\subsection{Quasiparticle distribution}
\comment{Brief presentation of the content of the subsection.}
The nonequilibrium current and the correlation of its fluctuations depends on the statistical distribution of the quasiparticle occupation.
We study classical dynamics of the occupation numbers of quasiparticle states, neglecting any coherent phenomena.
In other words, we only consider the evolution of the diagonal part of the density matrix in the basis of Fock states.
This neglects coherent many-particle interaction and therefore neglects the $8\pi$-periodic Josephson effect.
The non-adiabatic effects suppress the $8\pi$-periodic Josephson effect, and they have a larger rate in long junctions~\cite{Zhang2014}.
On the other hand, the $4\pi$-periodic Josephson effect becomes more pronounced in long junctions, justifying our approximation.
The dynamics of the junction is then determined by a rate equation which models possible relaxation processes.
In this section, we derive the quasiparticle distribution in long junctions with $2N$ levels and the rate equation governing its time evolution.

\begin{figure}[t]
\includegraphics[width=0.45\textwidth]{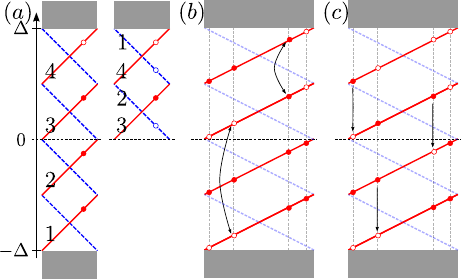}
\caption{(Color online) Schematic representation of energy eigenstates and their occupation in an ideal model of a four-level junction with a linear spectrum.
On the $x$ axis, the superconducting phase difference $\varphi$ varies always in the first Brillouin zone with $\varphi\in (-\pi,\pi]$.
Solid dots represent a particle occupying a level, while an empty circle, an unfilled level.
Panel (a) shows a comparison between two equivalent ways to count the states.
Left side shows the convention used in this paper, where only right-moving states are counted in order from $-\Delta$ to $\Delta$.
The right side shows the usual representation considering only positive-energy excitations, where it is necessary to consider both left- and right-moving states.
Note that a negative filled (empty) right-moving state corresponds in the usual picture to a empty (filled) left-moving state.
Panel (b) represents energetically favorable \textit{spin-flip} two-particle dissipation events where the system relaxes to the ground state from an initial excited state.
Note that the first process is equivalent in the alternative picture to a relaxation from a left mover to a right mover.
Panel (c) represents energetically favorable \textit{spin-conserving} relaxation events where, starting from the same initial quasiparticle distribution as in (b), the system relaxes to the ground state.
Note that the last process depicts a two-particle relaxation where a pair of counter-propagating quasiparticles are lost to the condensate.
Since (c) are faster processes which relax the system before (b), this initial distribution of quasiparticles is equivalent to the ground state for the rate equation~\eqref{approx_rate}.
In contrast, the quasiparticle distribution in $(a)$ is immune to spin-conserving relaxation processes.}
\label{fig:state_rep}
\end{figure}

\comment{Convention for counting the states in the system.}
Due to the particle-hole symmetry of the BdG Hamiltonians, every positive-energy eigenstate has a partner at opposite energy.
Nevertheless, a level and its opposite-energy partner [shown in Fig.~\ref{fig:state_rep} (a)] describe the same physical excitation.
Hence, a filled positive level is the same as an empty level at the opposite energy, and vice versa.
Therefore, the system has $4N$ eigenstates~\eqref{eigs} between $-\Delta$ and $\Delta$, but only $2N$ distinguishable quasiparticle excitations.
This leads to a total of $2^{2N}$ possible states describing the occupation of the Andreev levels in the junction at a certain time.
Since elastic back-scattering is not allowed, the level crossings in Fig.~\ref{fig:state_rep} are protected.
This allows us to identify $\varepsilon^+$ levels as carrying positive current (right moving) because $\partial_\varphi \varepsilon^+_i>0$ and $\varepsilon^-$ levels as carrying negative current (left moving).

A common way of counting the many-body states is to consider quasiparticle occupation only at positive energy, with both
right- and left-moving eigenstates.
We use a different convention where only right-moving eigenstates are considered, but at both positive and negative energy.
Therefore, an empty right-moving negative-energy state, represents physically a counter-propagating (left-moving) quasiparticle.
The levels are labeled in the order of increasing energy from the first level near $-\Delta$ to $2N$-th level near $\Delta$, half of the levels with positive energy and half with negative.

Since, in every period, a new eigenstate enters at $-\Delta$ and one leaves at $\Delta$, we relabel the levels in each period to always start from one.
To simplify the notation, we omit the superscript for the right-moving level energies, such that from now on $\varepsilon_i\equiv\varepsilon_i^+$.
Therefore, a system state $s$ is represented by a set of right-moving level occupation numbers:
\begin{equation}
s=\{s_1, s_2, \dots, s_{2N}\},
\end{equation}
with $s_j$ being the fermionic occupation number of Andreev level $j$, $s_j=0$ or 1.
The ground state has all negative energy levels filled and all the positive energy levels empty.

We consider a constant voltage $V$ between the superconducting leads turned on abruptly at $t=0$ such that $\varphi(t)=2eVt/\hbar+\varphi_0$ and the junction starts in equilibrium with no quasiparticle excitations.
Without loss of generality, we set the arbitrary initial phase difference between superconductors $\varphi_0=-\pi$, such that energy levels $\varepsilon$~\eqref{eigs} cross the Fermi level $E=0$ at times $t_n$ multiples of the driving period: $t_n=nT$ or $2\pi n/\omega_J$, with the Josephson angular frequency $\omega_J=2eV/\hbar$.
Since the spectrum is $2\pi$ periodic with the phase $\varphi$ and $\partial_\varphi\varepsilon_i>0$, one quasiparticle is added in the beginning of every period $T$.
In the absence of additional inelastic scattering, all $2N$ levels in the junction become filled after a time $NT$.
After all the levels are occupied, one new quasiparticle is excited at the Fermi level in every period, while the quasiparticle in the highest level escapes to the continuum spectrum at $E>\Delta$.
Since the pattern of quasiparticle occupation repeats when the phase varies by $2\pi$, the resulting current is $2\pi$ periodic and the usual integer Josephson effect ensues.

In contrast, inelastic scattering processes allow quasiparticles to annihilate, leading to partially occupied levels.
We classify them into spin-conserving and spin-flip dissipative processes (see Fig.~\ref{fig:state_rep}), which we expect to be fast and slow, respectively.
The spin-conserving processes include (a) single-particle relaxation of a quasiparticle into an energetically lower empty co-propagating state and (b) two-particle relaxation of a pair of two counter-propagating quasiparticles into the condensate.
In contrast, the spin-flip processes include (a) single-particle relaxation of a quasiparticle into an empty anti-propagating eigenstate of lower energy and (b) pairwise annihilation of co-propagating quasiparticles into the condensate.
We note here that even in the presence of relatively large Rashba spin-orbit coupling, the association of a pseudo-spin with the variable $s$ allows the nearly spin-conserving limit to be applicable.

The spin-conserving relaxation preserves the $2\pi$ periodicity of the Josephson current.
In absence of the spin-flip scattering, the bulk of the system has a quantized spin Hall conductance, and therefore injects a single spin $1/2$ into the junction every time the flux is increased by a flux quantum.
This excites a right-moving  Andreev bound state in the junction.
Eventually all the $2N$ levels of the junction fill up, following which the spin accumulated in each cycle is ejected from the junction into the bulk of the superconductor.
The Andreev bound state occupation is then $2\pi$-periodic in $\varphi$, leading to a $2\pi$-periodic current.

In contrast, spin-flip processes may empty any two right-moving levels, prevent the population of all the junction states, and consequently, the ejection of quasiparticles into the continuum.
In this case, the fermionic parity is not constant in every period, since the periodic injection of a particle in the lowest level is not offset by the periodic ejection of the quasiparticle from the highest level into the continuum.
This leads to a non-$2\pi$-periodic Josephson current, whose signatures will be investigated in the following sections.

The quasiparticle occupation is described by the probabilities $p_s(t)$ for the occurrence of any state $s$ at time $t$.
The rate equations model relaxation events in the junction, described by a time, energy, and state-dependent transition rate $\Gamma_{s\to s'}(t)$ from a state $s$ to $s'$.
The time evolution of the quasiparticle distribution is given by the rate equation:
\begin{subequations}\label{full_rate}
\begin{equation}\label{rate}
\frac{dp_{s}(t)}{dt}
=\sum_{s'}\Gamma_{s'\to s}(t) p_{s'}(t)-\sum_{s'} \Gamma_{s\to s'}(t) p_s(t),
\end{equation}
\begin{eqnarray}\label{rate_exp}
\Gamma_{s\to s'}&=&\sum_{1\le j<i}^{2N}
\Big\{\gamma_{ij}(t)\Big[s_is_j(1-s_i')(1-s_j')\Theta(\varepsilon_i+\varepsilon_j)
\notag\\
&&+(1-s_i)(1-s_j)s_i's_j'\Theta(-\varepsilon_i-\varepsilon_j)\Big]\\
&&+\chi_{ij}s_i(1-s_j)s_j'(1-s_i')\Theta(\varepsilon_i-\varepsilon_j)
\Big\}
\prod_{k\ne i,j}\delta_{s_k^{\phantom\prime}s_k'},\notag
\end{eqnarray}
\end{subequations}
with $\delta_{ij}$ the Kronecker delta.
The microscopic rates $\chi$ govern the fast spin-conserving dissipative processes: the relaxation of a quasiparticle on a lower empty co-propagating level, when $\sgn(\varepsilon_i)=\sgn(\varepsilon_j)$, and annihilation of counter-propagating quasiparticles when $\sgn(\varepsilon_i)\ne\sgn(\varepsilon_j)$.

The spin-flip relaxation rates $\gamma_{ij}$ depend on the microscopic origins of dissipation.
We consider either phenomenological constant rates $\gamma_{ij} =\gamma$, or $\gamma_{ij}(t)=\alpha|\varepsilon_i(t)+\varepsilon_j(t)|^3$ appropriate for coupling to a photon bath (see Appendix~\ref{app:circ}), with $\alpha$ the dissipation strength.
When $\sgn(\varepsilon_i)=\sgn(\varepsilon_j)$, they denote (a) the annihilation of two co-propagating quasiparticles and (b), when $\sgn(\varepsilon_i)\ne\sgn(\varepsilon_j)$, relaxation of a quasiparticle into a lower empty counter-propagating level.

In addition to the relaxation processes, at every $nT$ a new quasiparticle is excited in the junction, the lowest state becomes filled, and the quasiparticle occupations shift by one.
If the highest level $\varepsilon_{2N}$ near $E=\Delta$ is filled, the respective quasiparticle is lost to the continuum.
Therefore, the state probability $p_s(t)$ in Eq.~\eqref{rate}, satisfies boundary conditions:
\begin{eqnarray}\label{w}
p_s(nT+0^+)&=&\sum_{s'} W_{s'\to s}p_{s'}(nT-0^+),\\ W_{s'\to s}&=&s_1\prod_{j=1}^{2N-1}
\delta_{s_j'^{\phantom\prime}s_{j+1}}.\notag
\end{eqnarray}
Here, $W$ is a shift operator of the level occupation numbers.

For brevity, we rewrite Eq.~\eqref{rate} in vector form:
\begin{equation}
\frac{d\bm p(t)}{dt}=\bm\Gamma(t)\cdot\bm p(t),
\end{equation}
with $\bm p$ the $2^{2N}$-dimensional vector of state probabilities and $t\in(n,n+1)T$.
The corresponding evolution of the probability over one period is $\bm p(t+T)=U(t+T, t)\bm p(t)$ with time-evolution operator:
\begin{equation}\label{evo}
U(t+T, t)=\mathcal T e^{\int_{0}^{T-t}\bm\Gamma(t')dt'}W
\mathcal T e^{\int_t^{T}\bm\Gamma(t')dt'},
\end{equation}
and $\mathcal T$ denoting time-ordered product of operators.
The periodicity of the dissipation matrix  $\bm\Gamma(t+T)=\bm\Gamma(t)$ allowed us to bring all integrals in the first period $(0, T]$.

The periodic steady-state probability $\bm p_\infty(t+T)=\bm p_\infty(t)$ follows as a normalized solution to
\begin{equation}\label{steadyp}
[\bm 1-U(t+T,t)]\cdot\bm p_\infty(t)=\bm 0,
\end{equation}
with $\bm 0$ and $\bm 1$ the zero and the identity matrices, respectively.
Since $U(t+T,t)$ is a Markov matrix, it has always at least one steady-state solution.
Moreover, all states are either part of a single closed set of communicating states, or transient states towards this set.
\footnote{When spin-flip processes are neglected $\gamma=0$, the closed set contains a single state, the one with all levels filled.}
The steady state is unique, since the closed set has a unique steady state under Perron-Frobenius theorem.~\cite{Gantmacher2000}

\subsection{The fast relaxation approximation}
\label{subsec:fast_approx}
The rate equation~\eqref{full_rate} together with the boundary condition~\eqref{w} describes the evolution of the quasiparticle distribution in a $2N$-level junction in a space of $2^{2N}$ states.
The accessible state space and the rate equation simplifies in the limit when the spin-conserving relaxation is much faster than the spin-flip scattering, i.e., $\chi\gg \gamma$.
In this regime the system relaxes over the time scale $1/\chi$ to the lowest-energy state with a given total spin (i.e.,~the difference between the number of occupied positive levels and empty negative levels): when all the levels below a certain energy are occupied and the levels above are empty.
The slower spin-flip relaxation processes then reduce the total spin by removing a pair of quasiparticles, by annihilating a pair of positive levels or by creating a pair of occupied negative levels, followed by the quick relaxation to the lowest-energy state.
Therefore, except for the time fraction $O(\gamma/\chi)$ the system occupies one of the $2N+1$ lowest energy states with a fixed total spin and total number of particles $n$: $n\in\{0,1,\dots,2N\}$.
Consequently, the time evolution on the long-time scale is obtained by solving the rate equation for $p_n(t)$ in this reduced space.
Finally, note that while spin-conserving relaxation cannot generate non-$2\pi$-current signatures, it enhances the fractional Josephson signatures by keeping the system in the lowest-energy state with a given particle number, and therefore preventing excited quasiparticles from reaching continuum before $n = 2N$ (see Appendix~\ref{app:full_rate}).

The transition rate from state $n$ to $n'$ is the sum of all the transition rates to intermediate states that are accessible through a spin-nonconserving relaxation process:
\begin{eqnarray}\label{approx_rate}
\Gamma_{n\to n'}(t)
&=&\sum_{N+\Pi(\frac{t}{T})\le i<j}^{n}\gamma_{ij}(t)\delta_{n-2,n'}\notag\\
&&+\sum_{n< i<j}^{N+1-\Pi(\frac{t}{T})}
\gamma_{ij}(t)\delta_{n+2,n'}.
\end{eqnarray}
Here, if the lower bound of the sum is higher than its upper bound, the sum equals to zero and $\Pi(x) \equiv \Theta(\textrm{frac}(x) - 1/2)$, with $\textrm{frac}(x)$ the fractional part of $x$.
The first term in Eq.~(\ref{approx_rate}) is the loss of two occupied levels, energetically favorable when $n > N+1$ in the first half of a period and $n>N$ in its second half.
The second term models the gain of two occupied levels, favorable when $n<N$ in the first half of the period and $n<N-1$ in the second half of the period.
Note that the junction ground state $n=N$ remains always an absorbing state (immune to spin-flip relaxation processes), while additionally the excited state $n=N+1$ is an absorbing state in the first half of the period, and $n=N-1$ is an absorbing state in the second half of the period.
Finally, the shift operator $W$ in Eq.~\eqref{w} becomes in the reduced basis $W_{n\to n'}=\delta_{n+1, \min\{n',2N\}}$.
For a positive bias voltage, the state space could be further reduced by eliminating the transient states $0 \leq n < N-1$.
The remaining $N+2$ states are all communicating and form an irreducible Markov chain.

\subsection{Current and power spectrum}
The Josephson current $I$ carried by the junction consists of the ground-state contribution $I_{gs}$, and the nonequilibrium part $I_{ne}$, due to excited quasiparticle states:
\begin{equation}\label{Itot}
I=I_{gs}+I_{ne}.
\end{equation}
In the following, we consider a long Josephson junction with $2N$ levels.
Because each Andreev level carries current $i_0$ and there are $N$ levels filled in equilibrium, the nonequilibrium current equals
\begin{equation}\label{curr0}
 I_{ne}(t)=i_0  (n_s-N),
\end{equation}
with the total number of particles $n_s=\sum_{j=1}^{2N}s_j$.

In the steady state, $\langle I^\infty_{ne}(t)\rangle$ is $2\pi$ periodic (here and later $\langle x \rangle$ is the statistical average), and the approximate $4\pi$ periodicity manifests as a peak in the noise power spectrum at half-integer multiples of the Josephson frequency.~\cite{Fu2009,Houzet2013}
The finite-frequency power spectrum of the Josephson current reads as
\begin{equation}
P(\omega)=\lim_{C\to\infty}\frac{1}{C}
\int_0^C dt \int^C_0 dt'\langle I(t) I(t') \rangle e^{i\omega(t-t')}.
\end{equation}
Using the $2\pi$ periodicity of $p_s(t)$ in the steady state, the power spectrum simplifies to
\begin{equation}\label{pow1}
P(\omega)=\frac{1}{T}\int_0^{T} dt \int^\infty_0 dt'\langle I(t) I(t') \rangle e^{i\omega(t-t')}.
\end{equation}
When expanding the current operator using Eq.~\eqref{Itot}, the power spectrum splits into three contributions involving the correlators
$\langle I_{ne}I_{ne}\rangle$,
$\langle I_{gs}I_{gs}\rangle$, and
$\langle I_{ne}I_{gs}\rangle$.
Accordingly, the power spectrum decomposes into contributions from the
respective correlators:
\begin{equation}\label{pow}
P(\omega)=P_{ne\text{-}ne}(\omega)+P_{gs\text{-}gs}(\omega)+P_{ne\text{-}gs}(\omega),
\end{equation}
with the contribution from both $ne$-$gs$ and $gs$-$ne$ correlators included in the last term.

The terms in the power spectrum decomposition containing the contribution from the $2\pi$-periodic ground-state current  do not exhibit signatures of a fractional Josephson effect.
For example, $P_{gs\text{-}gs}(\omega)$ consists of a series of delta peaks at integer multiples of the Josephson frequency.
By substituting $I_{gs}$ from Eq.~\eqref{igs} in $P_{gs\text{-}gs}(\omega)$, it follows that the power spectrum at positive frequency reads as
\begin{equation}
P_{gs\text{-}gs}(\omega)=\frac{i_0^2}{2\pi}\sum_{k=1}^\infty
\frac{1}{k^2}
\delta(\omega-k\omega_J).
\end{equation}
The same holds for the cross-term contribution to the power spectrum since in the long-time limit the steady-state nonequilibrium current is independent of the ground-state current:
\begin{eqnarray}
P_{ne\text{-}gs}(\omega)
&=&\frac{2}{C}\textrm{Re}\int_0^C dt\int_0^C
dt'
\langle I^\infty_{ne}(t)\rangle
I_{gs}(t')
e^{i\omega(t-t')},\notag\\
&=&2\textrm{Re}\big[
\langle I^\infty_{ne}(\omega)\rangle
I_{gs}^*(\omega)\big].
\end{eqnarray}
Since both $\langle I^\infty_{ne}(\omega)\rangle$ and $I_{gs}(t)$ are $2\pi$ periodic, $P_{ne\text{-}gs}(\omega)$ is also a series of Dirac delta functions at integer multiples of the Josephson frequency.

The non-$2\pi$-periodic contributions to the Josephson effect are due entirely to the nonequilibrium correlator $\langle I_{ne}I_{ne}\rangle$.
Using the definition $\eqref{curr0}$ it reads as
\begin{equation}
\langle I_{ne}(t+\tau)I_{ne}(t)\rangle
=i_0^2\sum_{s,s'}(n_{s'}-N)(n_s-N)p(s', t+\tau; s, t),
\end{equation}
where the joint probability $p(s', t+\tau; s, t)$ denotes the probability that the system is in state $s'$ at time $t+\tau$ ($\tau>0$) and in state $s$ at time $t$.
The joint probability is further expanded using the conditional probability $p(s', t+\tau; s, t)=p(s', t+\tau|s, t) p(s, t)$.
Since the quasiparticle occupation dynamics is Markovian, we compute the conditional probability $p(s',t+\tau|s,t)$ by solving Eqs.~\eqref{full_rate} or~\eqref{approx_rate} with the initial condition $p_s(t)=1$.
Furthermore, in the long-time limit, $t$ is far from an initial time $t_0$, such that the system has already reached its steady state and $p_s(t)$ may be replaced by $p_{s,\infty}(t)$.
Consequently, the power spectrum~\eqref{pow1} reads as
\begin{eqnarray}\label{pownene}
P_{ne\text{-}ne}(\omega)&=& 2i_0^2
\int_{0}^{\infty}d\tau \cos(\omega\tau)
\frac{1}{T}\int_0^T dt \sum_{s,s'}(n_{s'}-N)\notag\\
&&\times (n_s-N) p(s',t+\tau|s, t)p_{s,\infty}(t).
\end{eqnarray}
The expression~\eqref{pownene} allows us to compute the noise power spectrum by numerically determining the steady state $p_{s,\infty}(t)$, solving the rate equation with different initial conditions and numerical integration.

\section{Short junctions}
\label{sec:1lev}
In order to illustrate the role of two-particle relaxation in the appearance of the $4\pi$-periodic Josephson effect, we consider first a minimal setup for the case of short junctions where there are at most two levels in the junction.
In the short-junction limit $E_T\gg\Delta$, any terms on the order of $\Delta/E_T$ are neglected.
Consequently, the dispersion has a cosine shape~\eqref{sj} with a single level in the junction, and therefore no two-particle relaxation for most values of $\varphi$.
Nevertheless, for any finite ratio $\Delta/E_T$, there are always two levels in the junctions near $\varphi=2n\pi$ allowing for two-particle relaxation.
The small phase interval over which the two levels coexist reads as, from Eq.~\eqref{spect}, $\Delta\varphi\approx 4\Delta/E_T$.

Since the spectrum is $2\pi$ periodic, it is sufficient in the following to focus on a single period $\varphi\in(-\pi,\pi]$.
The two right-moving states coexisting at $\varphi\simeq 0$ are determined from Eq.~\eqref{spect}:  $\varepsilon_0=\Delta\cos(\varepsilon_0/E_T-\varphi/2)>0$ and $\varepsilon_1=-\Delta\cos(\varepsilon_1/E_T-\varphi/2)<0$.
At $\varphi=0$, the negative-energy state $\varepsilon_1$ is empty, which is equivalent to having an excited left-moving quasiparticle in eigenstate $\varepsilon^-_0$.
Relaxation of the right-moving quasiparticle into an empty left-moving quasiparticle or equivalently emptying levels $\varepsilon_0$ and $\varepsilon_1$ leads to an energy change:
\begin{equation}
-(\varepsilon_0+\varepsilon_1)\approx -\frac{2\Delta^2}{E_T}\sin(\varphi/2).
\end{equation}
Therefore, two-particle relaxation is energetically favorable when $\varepsilon_0>-\varepsilon_1$ for $0<\varphi<2\Delta/E_T$.
This conclusion holds in general due to spectrum periodicity, such that two-particle relaxation is allowed whenever $2n\pi<\varphi<2n\pi+2\Delta/E_T$.

\begin{figure}[t]
\includegraphics[width=0.72\columnwidth]{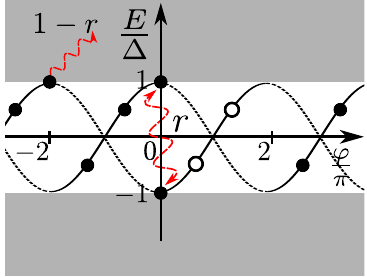}
\caption{Model for short junctions $E_T\gg\Delta$.
The figure shows two-particle relaxation generating a $4\pi$-periodic occupation of single right-moving state.
The sine-shaped curves are the Andreev state energies as a function of $\varphi$.
The overlap of two energy levels near $\varphi=2\pi n$ is not shown.
By convention we only consider right-moving states (black solid lines) and their occupation marked with solid dots for filled and open circles for empty states.
At $\varphi=-2\pi$ the excited quasiparticle escapes into the continuum (gray area) with probability $1-r$.
Therefore, the right-moving state remains filled for $-2\pi < \varphi < 0$.
In contrast, at $\varphi=0$ a two-particle relaxation process takes place (with probability $r$).
Then, the Andreev state becomes empty for $0 < \varphi < 2 \pi$.
Consequently, the occupation of the state in $(-2\pi,0)$ is recovered only after two periods for $2\pi < \varphi < 4\pi$.
\label{fig:toy_draw}
}
\end{figure}

When an excited quasiparticle is close to the continuum at any $\varphi\in [2\pi n,2\pi n+2\Delta/E_T)$, it can either go through a two-particle relaxation process with a probability $r$, or escape into the continuum with probability $1-r$.
Since for short junctions $\Delta/E_T\ll 1$, we model both relaxation processes as occurring at discrete times when $\varphi=2\pi n$ (see Fig.~\ref{fig:toy_draw}).
After this simplification, the effect of two-particle relaxation becomes formally equivalent to the opening of the spectral gap by an applied in-plane magnetic field~\cite{Houzet2013, Badiane2013}.
In that case, $1-r$ is the probability that the fermion parity of the junction changes due to Landau-Zener tunneling across a magnetically induced gap at $\varphi=2\pi n$.
Because the models are identical, we naturally reproduce the results of Refs.~\onlinecite{Houzet2013, Badiane2013} in the short-junction limit.

We choose $\varphi_0=0$, so that $\varphi=2eVt/\hbar$, and the occupation probability of the single Andreev level is constant within each period $p(t) \equiv p(n)$ with $n=\lfloor t/T\rfloor$.
The master equation now assumes the form:
\begin{equation}
p(n)=1-rp(n-1).
\end{equation}
In the limit of infinitely strong two-particle dissipation $r=1$, the occupation probability has period $2T$ and the level occupation alternates indefinitely.
Without two-particle dissipation $r=0$, a steady state where the Andreev level is always filled $p(n)=1$ is reached already after a single period.

The conditional probability of the level to be filled after any $k$ periods reads as:
\begin{equation}\label{condprob}
p(n+k)=\frac{1}{1+r}
+(-r)^k\left[p(n)
-\frac{1}{1+r}
\right],
\end{equation}
and accordingly the steady-state occupation probability follows in the limit $k\to\infty$:
\begin{equation}
p_\infty=\frac{1}{1+r}.
\end{equation}

The current associated with left- or right-moving eigenstates in an arbitrary period $k$ follows from the dispersion Eq.~\eqref{sj}:
\begin{equation}
I_k^{\pm}(\varphi)\approx\pm I_c |\sin(\varphi/2)|,\quad
I_c=\frac{e\Delta}{2\hbar},
\end{equation}
with $I_c$ the critical current in the short-junction limit and $\varphi\in 2\pi(k,k+1]$.
We have neglected small corrections to the dispersion on the order of $\Delta^2/E_T$ near the continuum at $\pm\Delta$.

The mean current in the $k$-th period reads as
\begin{equation}
\langle I_k(\varphi)\rangle
= I_c(2p_k-1)|\sin(\varphi/2)|,
\end{equation}
leading to a $2\pi$-periodic average steady-state current:
\begin{equation}
\langle I_\infty(\varphi)\rangle=I_c\frac{1-r}{1+r}|\sin(\varphi/2)|.
\end{equation}
Therefore, the dc current obtained by averaging $\langle I_\infty(\varphi)\rangle$ over $\varphi$ reads as
\begin{equation}
I_{\rm dc}=\frac{2I_c}{\pi}\frac{1-r}{1+r}.
\end{equation}
As expected, the dc current decreases to zero in the limit of strong dissipation $r\to 1$.

\begin{figure}[t]
\includegraphics[width=\columnwidth]{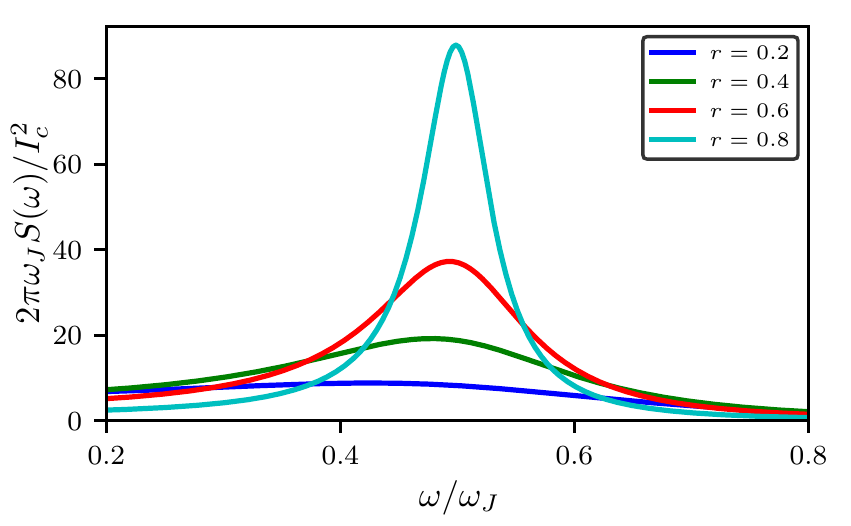}
\caption{(Color online) Peak at half the Josephson frequency in the noise spectrum of the supercurrent in short-junctions [Eq.~\eqref{toynoise}] for different two-particle dissipation probabilities $r$.}
\label{fig:1noise}
\end{figure}

The power spectrum from Eq.~\eqref{pow1} is determined using the autocorrelation function for $t>t'$:
\begin{eqnarray}
\langle I(t)I(t')\rangle
&=& I_c^2\left[\frac{(1-r)^2}{(1+r)^2}
+\frac{
4r(-r)^{\lfloor \frac t T\rfloor-\lfloor \frac{t'}{T}\rfloor}}{(1+r)^2}\right]\notag\\
&\times&\left|\sin\left(\frac{\pi t}{T}\right)\sin\left(\frac{\pi t'}{T}\right)\right|.
\end{eqnarray}
Here, the first term is the product of mean currents in the long-time limit $\langle I_\infty(t)\rangle\langle I_\infty(t')\rangle$.
Since these mean currents are $2\pi$ periodic, they yield delta peaks in the power spectrum at integer frequencies.
In the following, we focus on the non-trivial part of the spectrum and investigate the noise spectrum, $S(\omega)=P(\omega)-|\langle I_\infty(\omega)\rangle|^2$.
Integration over the autocorrelator in Eq.~\eqref{pow1} yields
\begin{equation}\label{toynoise}
S(\omega)
=\frac{1-r}{1+r}\frac{I_c^2}{2\pi\omega_J}
\frac{1}{(\frac1 4-\frac{\omega^2}{\omega_J^2})^2}
\frac{4r\cos^2(\frac{\pi\omega}{\omega_J})}{(1-r)^2+4r\cos^2(\frac{\pi\omega}{\omega_J})}.
\end{equation}
As expected, Eq.~\eqref{toynoise} recovers the functional form of the noise spectrum from Ref.~\onlinecite{Houzet2013}.
A peak in $S(\omega)$ at $\omega_J/2$ appears for strong two-particle relaxation $1-r\ll 1$ (see Fig.~\ref{fig:1noise}).
In this limit, the peak has a Lorentzian shape with the height $\pi I_c^2/(1-r)\omega_J$.
The width at half-maximum gives the inverse lifetime of the $4\pi$-periodic mean current: $(1-r)\omega_J/\pi=(1-r)/T$, and it matches the parity lifetime $\tau_{4\pi}$ predicted by Eq.~\eqref{condprob}:
\begin{equation}\label{sjtau4pi}
\tau_{4\pi}=\frac{T}{1-r}\approx -\frac{T}{\ln r}.
\end{equation}

We have therefore shown that also in time-reversal symmetric short Josephson junctions, two-particle relaxation can create a $4\pi$-periodic ac Josephson effect.
Nevertheless, we expect this effect to be suppressed with the junction size because the probability of two-particle relaxation $r\propto \Delta/E_T\ll 1$.
Instead, we will focus in the following on long Josephson junctions, where the case for two-particle relaxation as a source for observable $4\pi$ periodicity becomes stronger.
This is due to the existence of many subgap levels, such that there are more channels for relaxation, and spin-flip dissipation processes may occur in general at arbitrary phase values.

\section{Long junctions}
\label{sec:Nlev}
\subsection{Introduction and asymptotic behavior}
We now turn to analyze long Josephson junctions with multiple Andreev levels $2N\simeq 2\Delta/\pi E_T$ and a linear dispersion relation~\eqref{eigs}.
The subsequent rate equation describing the dynamics of the $2^{2N}$ vector of state probabilities can no longer be solved analytically.
Using the methods described in Sec.~\ref{sec:model}, we identify signatures of the fractional Josephson effect mainly through numerical simulations and asymptotic analysis.
We focus on the fast relaxation approximation~(Sec.~\ref{subsec:fast_approx}) where the system evolves in a reduced $2N+1$ set of states.
The power spectrum governed by the full rate equation~\eqref{full_rate} provides qualitatively similar results as we show in Appendix~\ref{app:full_rate}.

Simulations start with zero excited quasiparticles in the junctions and an initial phase difference between superconductors $\varphi_0=-\pi$.
Therefore, at initial time $t=0$, the ground-state current is at its minimum $I_{gs}=-i_0/2$.
The first Andreev level crosses the Fermi level after a period $T$ and it carries an excited quasiparticle, thus contributing to the nonequilibrium current $I_{ne}$.
The time-evolution of the system is solved through numerically propagating the vector of probabilities from the initial state.

Before eventually reaching a periodic steady state, i.e.,~when the mean current becomes $2\pi$ periodic, the system goes through a transient regime.
Two time scales define the evolution in the transient regime.
The first one is set by the time required to fill the $N$ energy levels of the junction in the absence of dissipation.
Due to injection of a particle every period, this time scale is $\tau_\textrm{fill}=NT$.
Simulations are required to exceed $\tau_\textrm{fill}$.
The second time scale $\tau_{4\pi}$ is characteristic for the decay of the $4\pi$-periodic mean current.
This may be extremely long in our ideal setup, growing exponentially (as we will establish later) with the number of levels and dissipation strength.
Nevertheless, Eq.~\eqref{steadyp} determines the steady state even when $\tau_{4\pi}$ exceeds feasible simulation times.
In fact, the long correlation time $\tau_{4\pi}\gg T$ will be shown to be the regime where a sharp fractional peak develops in the power spectrum.
Asymptotic analysis uncovers in the following the scaling behavior for $\tau_{4\pi}$, confirmed by simulations in the next subsection.

As explained in Sec.~\ref{sec:intro}, the condition to have a clear signature for the $4\pi$-periodic effect is that two-particle dissipation is effective enough such that quasiparticles have a small probability to reach the continuum.
Since the $4\pi$-periodic signal is due to spin-flip relaxation processes, we consider in the following the evolution of the system in the reduced space of $2N+1$ steady states of spin-conserving relaxation processes~(see Sec.~\ref{subsec:fast_approx}).
Let $n \ge N$ be the average steady-state occupation of a $2N$-level junction, or in other words that there are $\approx n-N$ right-moving excited quasiparticles.
In this $2\pi$-periodic state, the generation of a quasiparticle at the lowest level is compensated by the loss due to two-particle dissipation.
Since there are on average $(n-N)(n-N+1)/2$ pairs of quasiparticles which may annihilate within a period $T$ with rate $\gamma$, the total loss of quasiparticles in the case of time-independent dissipation reads as
\begin{equation}\label{gn1}
\gamma (n-N)^2T\sim 1.
\end{equation}
This condition translates  to an average number of filled levels
\begin{equation}
n\sim N+(\gamma T)^{-1/2}.\label{eq:navg}
\end{equation}
The above expression assumes a long junction with many levels and excited quasiparticles $2N>n-N\gg 1$.

If there are many levels, it is necessary to have a large number of excited quasiparticles before one quasiparticle is ejected into the continuum.
In the absence of such a process, the fermion parity is only changed by the injection of quasiparticles from negative energies with every $2\pi$ variation in $\varphi$. In the case of fermion parity preserving spin-flip relaxations, the $4\pi$-periodic oscillations of fermion parity leads to a $4\pi$-periodic current.
However, the above argument misses rare events where a string of relaxation events do not occur.
Including such processes can lead to the ejection of a quasiparticle into the continuum.
Such events flip the $4\pi$-periodic current reducing the periodicity of the current to $2\pi$ beyond a potentially long but finite correlation time.
In order to estimate this correlation time, we compute the probability that the system evolves from the steady state $n$ to the state with all levels occupied, such that the highest quasiparticle subsequently escapes into the continuum.
The shortest path to the state $2N$ requires the system to advance in increasing order over states $n, n+1,\ldots, 2N$.
A further simplification to the rate equation~\eqref{approx_rate} involves neglecting the coupling between the differential equations for different states $j$.
We see that this approximation is equivalent to assuming $p_{j+2}\ll p_j$, which is the case for larger dissipation rates, leading to lifetimes $\tau_{4\pi}$ longer than a period $T$.
Since this approximation underestimates $p_j$ (by ignoring decay of states into $j$), the scaling analysis gives a lower bound for the probability to eject a particle into continuum, and correspondingly an upper bound for the lifetime $\tau_{4\pi}$.
Keeping in mind that within each period the state $j$ evolves into the state $j+1$, the approximations discussed above yield that the solution to Eq.~\eqref{rate} can be approximated as:
\begin{equation}
p_j \sim p_{j-1} \exp\bigg[-\gamma\pmat{j-N\\2} T\bigg],
\end{equation}
with the binomial coefficient being a result of counting the number of possible ways in which two particles may be lost due to dissipation
in Eq.~\eqref{approx_rate}.

\begin{figure}[t]
\includegraphics[width=\columnwidth]{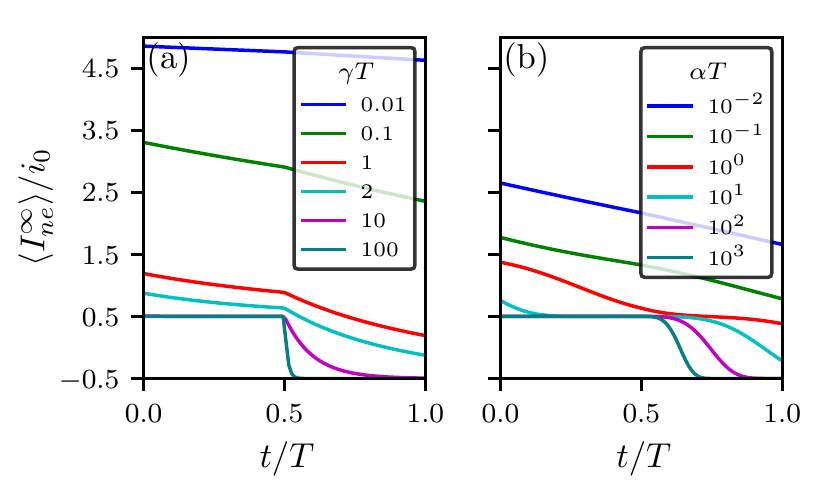}
\caption{(Color online) Mean periodic nonequilibrium steady-state current for an $2N=10$ level junction at different dissipation rates. The rates in the legend are in units of Josephson frequency $1/T$ either for (a) time and energy-independent relaxation ($\gamma T$) or (b) time and energy-dependent relaxation [$\alpha T$, with dissipation strength $\alpha$ in units of $(\pi E_T)^3$].
At small dissipation rates $\alpha T\ll 1$ or $\gamma T\ll 1$, all positive-energy levels in the junction are occupied and contribute to a current $\approx Ni_0$.
}
\label{fig:mean}
\end{figure}

If the system reaches the state $j=2N$, then it ejects a quasiparticle at the end of a period.
Therefore, the average time over which a quasiparticle is emitted, $\tau_{4\pi}$, relates to the inverse of the
probability that the system is in state $j=2N$ (excited here from the steady state $n$):
\begin{eqnarray}\label{tauprob}
\frac{T}{\tau_{4\pi}}\sim p_{j=2N}
\sim \exp\bigg[-\gamma\sum_{j=n}^{2N}\pmat{j-N\\2}T\bigg].
\end{eqnarray}
This leads to an estimate for the correlation time:
\begin{equation}\label{inv1}
\tau_{4\pi}\sim T\exp[(\gamma (n-N)^2(2N-n) T)].
\end{equation}
When $\tau_{4\pi} \le T$ a quasiparticle is ejected almost every period, which means that the $4\pi$-periodic component of the current flips almost every period and is therefore ill-defined.
Thus $\tau_{4\pi} \sim T$ represents a critical value of dissipation below which the $4\pi$-periodic current disappears.
Therefore, Eqs.~\eqref{gn1} and~\eqref{inv1} with $\tau_{4\pi}\sim T$ allow us to estimate the critical dissipation rate $\gamma_c$ and number of occupied levels $n_c$:
\begin{equation}\label{gcrit}
\gamma_c\sim N^{-2},\quad n_c\sim 2N,
\end{equation}
that demarcates the appearance of a $4\pi$-periodic component of the current with a long correlation time for weak dissipation and large $N$.
In the regime of strong dissipation rates $\gamma\gg\omega_J$, any quasiparticle pair is annihilated within a period, so the system is in a steady state close to ground-state $n\simeq N$ (see also Fig.~\ref{fig:mean}).
This means that $n\simeq N$ and the estimate for $\tau_{4\pi}$ using Eq.~\eqref{tauprob} is revised to:
\begin{equation}\label{glarge}
\frac{\tau_{4\pi}}{T}
\sim \prod_{j=0}^N e^{\gamma (
\begin{smallmatrix}
j\\2
\end{smallmatrix}
) T
}
 \sim e^{\gamma N^3T},
\end{equation}
which clearly shows how the correlation-time $\tau_{4\pi}$ for the $4\pi$-periodic component of the current diverges exponentially as
the number of levels $N$ and Josephson period $T$ increases.

Similar arguments apply for the case of energy-dependent dissipation, but lead to different scaling behaviors for critical dissipation.
In the steady state, the excitation of one quasiparticle due to driving is compensated by the quasiparticle relaxation:
\begin{equation}\label{an1}
\alpha T\sum_{N<i<j}^{n}(i+j)^3\sim \alpha T (n-N)^5\sim 1.
\end{equation}
Moreover, the probability that a quasiparticle escapes from the steady state by advancing to the $2N$ state follows using the same reasoning leading to Eqs.~\eqref{tauprob} and~\eqref{inv1}:
\begin{equation}\label{tau2}
\frac{\tau_{4\pi}}{T}
\sim\exp\bigg[
\alpha \sum_{k=n}^{2N}
\sum_{N<i<j}^k(i+j)^3 T
\bigg]\sim e^{\alpha(n-N)^5 (2N-n)T}
.
\end{equation}
Therefore, the scaling of the critical dissipation strength $\alpha_c$ and the average number of quasiparticles $n_c$ follow from the estimate of Eq.~\eqref{an1} and the criticality condition $\tau_{4\pi}\sim T$ in Eq.~\eqref{tau2}:
\begin{equation}\label{acrit}
\alpha_c \sim N^{-5},\quad n_c\sim 2N.
\end{equation}
As in the case of time and energy-independent relaxation rates, the scaling arguments which assumed $n-N\gg 1$ are consistent with the results for long junctions with many Andreev levels $N\gg 1$, since $n_c\sim 2N$.

In the limit of strong dissipation, the average number of quasiparticles tends to $n\simeq N$ in the steady state, which is close to the ground-state distribution.
Therefore, the scaling law for the lifetime in the strong-dissipation regime reads as
\begin{equation}\label{alarge}
\tau_{4\pi}\sim T e^{\alpha N^6 T}.
\end{equation}
Note that the above relations hold for $\alpha$ in units of $(\pi E_T)^3$, used in simulations, while the physical dissipation strength $\alpha_\textrm{phys}=\alpha/(\pi E_T)^3\sim \alpha N^3$.
Therefore, the scaling of the critical strength reads as $\alpha_{c,\textrm{phys}}\sim N^{-2}$ and, for strong dissipation, $\tau_{4\pi}\sim T\exp(\alpha_\textrm{phys}N^3T)$.

In the strong-dissipation limit, the $4\pi$-periodic part of the current develops an exponentially long correlation 
time [Eqs.~\eqref{glarge} and~\eqref{alarge}] making the width of the peak in the power spectrum difficult to resolve within our simulation time.
Nevertheless, we observe in our simulations the expected asymptotic behavior even for relatively small dissipation strength ranges and number of levels.

The mean current in the steady state follows readily in the limit of strong dissipation.
Since energy levels in the long junction are linear in phase with a fixed slope, the mean current is related to the mean number of excited quasiparticles in the junction $i_0(\langle n\rangle-N$).
Simulations in Fig.~\ref{fig:mean} show that in the limit of strong dissipation the mean current tends to a step-function shape.
This result is readily understood from the rate equation~\eqref{approx_rate} by identifying the absorbing states of the Markov chain in each half of a period.
In the first part of the period there are two absorbing states, the ground state $N$ which carries zero current $I_{ne}=0$ and the state with one excited right-moving quasiparticle $N+1$, $I_{ne}=i_0$.
Since each state comes with a probability $1/2$ to be realized, the mean nonequilibrium current in the steady state, in the limit of strong dissipation is $0.5 i_0$.
Similarly, in the second half of the period, the absorbing states are the ground state $N$ and $N-1$ (physically the state with a left-moving quasiparticle), due to relaxation of a right-moving particle becoming energetically favorable.
Consequently, the mean nonequilibrium current is $-i_0/2$.
Indeed, the limiting behavior of the mean nonequilibrium current in Fig.~\ref{fig:mean} reads as
\begin{equation}\label{mean_strong}
\lim_{\gamma,\alpha\to\infty}\frac{\langle I_{ne}^\infty(t)
\rangle}{i_0}
=\left[\frac{1}{2} - \Pi\left(\frac{t}{T}\right)
\right].
\end{equation}

The dc current contribution of the Andreev levels in the junction is obtained by averaging the $2\pi$-periodic steady-state current over a period.
The resulting current-voltage characteristic is shown in Fig.~\ref{fig:dc} for both (a) time-independent and (b) time-dependent dissipation.
In the low-bias or strong-dissipation limit, the occupation essentially follows the ground state of the appropriate fermion parity, leading to an almost vanishing average current due to perfect compensation of mean currents inside a period~[Eq.~\eqref{mean_strong}].
However, due to excitation of a particle in every period, it is equally likely that a single positive level becomes occupied (i.e.,~the state $N+1$, with different fermion parity from the ground state).
In this case, according to the rate equation~\eqref{approx_rate}, when the quasiparticle is excited beyond the first (positive energy) crossing in the Andreev spectrum it is favorable for the state to decay into the state $N-1$ on a time-scale $\gamma^{-1}$ for time-independent dissipation.
Therefore, a straightforward calculation gives the average current over the period:
\begin{align}\label{eq:idc1}
I_{\rm dc}\simeq \frac{i_0}{\gamma T}
=\frac{2eV i_0}{h\gamma},
\end{align}
which linearly goes to zero at small bias voltages as seen from Fig.~\ref{fig:dc}(a).
In the limit of a very long junction $N\gg 1$, the average occupancy is given by Eq.~\eqref{eq:navg}.
This leads to a dc current at intermediate voltages where $\gamma T\ll N^2$ that is given by
\begin{align}\label{eq:idc2}
I_{\rm dc}\sim i_0(\gamma T)^{-1/2}\propto \sqrt{V},
\end{align}
which is non-linear in a rather $N$-independent way as seen from Fig.~\ref{fig:dc}(a).
Therefore, the observation of a linear voltage dependence [Eq.~\ref{eq:idc1}] or the square-root voltage dependence
[Eq.~\ref{eq:idc2}] of the dc current indicates a low filling of the junction.

In the small dissipation limit or large voltage limit (but $eV<\Delta$) relaxation becomes ineffective and all $2N$ levels in the junction are eventually occupied (physically only the $N$ right-moving excited quasiparticles survive).
This leads to a total dc current which saturates at $N i_0$.
For a large number of level junction, the saturation is difficult to observe since it requires exponentially small dissipation rates (see Fig.~\ref{fig:dc}).

At very strong dissipation rates, the simulations are unable to faithfully reproduce the exponentially narrow fractional peak in the power spectrum due to a limited frequency resolution.
Nevertheless, we determine analytically the qualitative features of the power spectrum in the asymptotic limit of strong dissipation, $\tau_{4\pi}\gg T$ or rates $\gamma\gg\omega_J$.
In the following, we prove that indeed the fractional Josephson peak in this parameter regime has a Lorentzian shape with height proportional to the lifetime $\tau_{4\pi}$.

The fermion parity $\sigma(t)=\pm 1$ in an ideal long QSH Josephson junction without two-particle dissipation is constant since the excitation of a quasiparticle at the Fermi level is offset by loss of a quasiparticle to continuum.
In contrast, strong two-particle relaxation may prevent quasiparticles to reach the continuum through recombination and loss of quasiparticles as soon as they are excited in the lower Andreev levels.
Hence, the fermion parity flips every $2\pi$ change of phase due only to the $2\pi$-periodic excitation of a quasiparticle at the Fermi level.
Since the fermion parity is recovered only after a $4\pi$ phase change, the Josephson current is $4\pi$ periodic.
In this limit, the fermion parity autocorrelator reads as
\begin{equation}
\langle \sigma(t)\sigma(t')\rangle
\approx \textrm{sgn}
[\cos(\pi t/T)\cos(\pi t'/T)]e^{-|t-t'|/\tau_{4\pi}}.
\end{equation}
Defects to the $4\pi$-periodic order occur on the scale of time intervals $|t-t'|$ longer than $\tau_{4\pi}$ due to a finite probability to promote quasiparticles to the last level and to eject them into the continuum.
Therefore, in the long-time limit, rare events ultimately decorrelate the current yielding $\langle\sigma(t)\rangle=0$ and a $2\pi$-periodic mean current ensues.

Signatures of $4\pi$ periodicity are still captured in the power spectrum of the junction.
We focus here only on the nonequilibrium current autocorrelator which yields the nontrivial signal.
The autocorrelator with explicit dependence on the fermion parity reads as
\begin{eqnarray}\label{autoco}
\langle I_{ne}(t)I_{ne}(t') \rangle&=&\sum_{\sigma\sigma'}
E[I_{ne}(t)I_{ne}(t')|\sigma(t)=\sigma,\sigma(t')
=\sigma']\notag\\
&&\times p(\sigma(t)=\sigma;\sigma(t')=\sigma'),
\end{eqnarray}
with $E[\dots|\dots]$, the conditional expected value.
Any topological character of the power spectrum must be related to the cases where the $4\pi$-periodic fermion parity lifetime $\tau_{4\pi}$ is long compared to fluctuations of the quasiparticle occupation $\tau_{qp}$ and the inverse Josephson frequency.
In the strong two-particle relaxation limit, we assume that the conditional expectation value of the current factorizes as:
\begin{equation}
E[I_{ne}(t)I_{ne}(t')|\sigma(t)\sigma(t')]
\approx E[I_{ne}(t)|\sigma(t)] E[I_{ne}(t')|\sigma(t')],
\end{equation}
for $|t-t'|\gg t_{qp}$.
This is because at times much longer than $\tau_{qp}$, aspects of the quasiparticle occupation apart from the fermion parity should become completely uncorrelated.

From the autocorrelator definition
\begin{equation}
\langle\sigma(t)\sigma(t')\rangle=\sum_{\sigma\sigma'}\sigma\sigma'p(\sigma(t)=\sigma;\sigma(t')=\sigma'),
\end{equation}
we obtain the joint probability distribution for the fermion parity in Eq.~\eqref{autoco}:
\begin{equation}
p(\sigma(t);\sigma(t'))=\frac{1+\sigma(t)\sigma(t')
\langle\sigma(t)\sigma(t')\rangle}{4}.
\end{equation}
Therefore, the current autocorrelator reduces to:
\begin{equation}\label{auto}
\langle I_{ne}(t)I_{ne}(t') \rangle
=\langle I_{ne}(t)\rangle
\langle I_{ne}(t')\rangle
+I_F(t)I_F(t')
\langle \sigma(t)\sigma(t')
\rangle,
\end{equation}
with the parity dependent average current:
\begin{equation}
I_F(t)=\frac{1}{2}\sum_{\sigma}\sigma E[I_{ne}(t)|\sigma(t)=\sigma].
\end{equation}

In the limit of a large $\tau_{4\pi}$, the leading contribution of the nonequilibrium current to the power spectrum reads:
\begin{eqnarray}
P_{ne\text{-}ne}(\omega)&\approx&
|\langle I_{ne}(\omega)\rangle|^2
+\int_0^{2T} dt dt'
I_F(t)I_F(t')\notag\\
&&\times\sum_n e^{i\omega(t-t'+2nT)}
\langle \sigma(t)\sigma(t')\rangle.
\end{eqnarray}
The first term depends on the product of $2\pi$-periodic mean currents and it only contributes to integer peaks in the power spectrum.
The second contribution to the power spectrum reads after performing a Poisson resummation over the correlator:
\begin{align}\label{powlarge}
&P_{ne\text{-}ne}(\omega)-|\langle I_{ne}(\omega)\rangle|^2 =\frac{\omega_J}{2\pi\tau_{4\pi}}\int_0^{2T} dtdt'\sum_k I_F(t)I_F(t')\notag\\ &\times\frac{e^{ik\omega_J(t-t')/2}}{(\omega-k\omega_J/2)^2+1/\tau_{4\pi}^2}\text{sgn}[\cos(\omega_J t/2)\cos(\omega_J t'/2)],\notag\\
&=\sum_k |\bar I_{k}|^2
\frac{\omega_J/2\pi\tau_{4\pi}}
{(\omega-k\omega_J/2)^2 +1/\tau_{4\pi}^2},
\end{align}
with $\bar I_{k}=\int^{2T}_0 dt I_F(t) \text{sgn}[\cos(\omega_Jt/2)]e^{ik\omega_J t/2}$.
The strong dissipation result Eq.~\eqref{powlarge} shows that the lifetime $\tau_{4\pi}$ is proportional to the peak height, which is  displayed in Figs.~\ref{fig:s},~\ref{fig:peak}, and~\ref{fig:powspec_td}, and it is inversely proportional to the peak width.

\subsection{Numerical results}
Following the qualitative analysis of the $4\pi$-periodic current behavior, we now discuss the results of our numerical simulations in $2N$-level junctions for both models of dissipation.

\begin{figure}[t]
\includegraphics[width=\columnwidth]{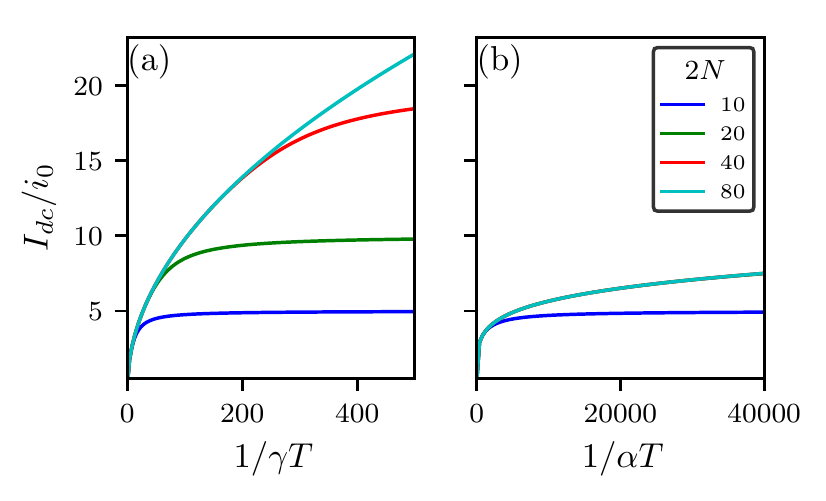}
\caption{(Color online) Current-voltage characteristic with increasing number of $2N$ levels in the junction.
Dissipation rates are (a) time-independent $\gamma$ or (b) time and energy-dependent, with dissipation strength $\alpha$ in units $(\pi E_T)^3$.
For strong dissipation or small voltage $V=h/2eT$ the dc current tends to 0, while at large voltage (but $eV<\Delta$) or low dissipation rates the dc current must saturate at $Ni_0$.
The panels share the legend showing the number of $2N$ levels in the junctions.
}
\label{fig:dc}
\end{figure}

The mean current in the steady state is $2\pi$ periodic, as determined by the state probability vector in the long-time limit~\eqref{steadyp}.
The computation of the steady-state vector additionally simplifies for the energy-independent dissipation rate model since the dissipation matrix $\bm\Gamma$ is time independent in each half of a period for the evaluation of Eq.~\eqref{evo}.
For time and energy-dependent dissipation, the state probability vector in the long-time limit~Eq.~\eqref{steadyp} is obtained through finite-time-difference evaluation of Eq.~\eqref{evo} and the resulting mean nonequilibrium current is shown in Fig.~\ref{fig:mean}(b).
As expected from Eq.~\eqref{mean_strong}, for both models, in the limit of strong dissipation $\gamma T\gg 1$ or $\alpha T\gg 1$, the mean number of quasiparticles in the junction relaxes towards the same distribution and yields the same mean current.
Integrating the steady-state current over a period yields the dc current shown in Fig.~\eqref{fig:dc}.
Both low and strong-dissipation limits discussed in the previous section are confirmed in the numerical simulations.
At low dissipation, the dc current saturates at $Ni_0$ and at strong dissipation it goes to linearly in voltage and inverse dissipation rate to zero.

\begin{figure}[t]
\includegraphics[width=\columnwidth]{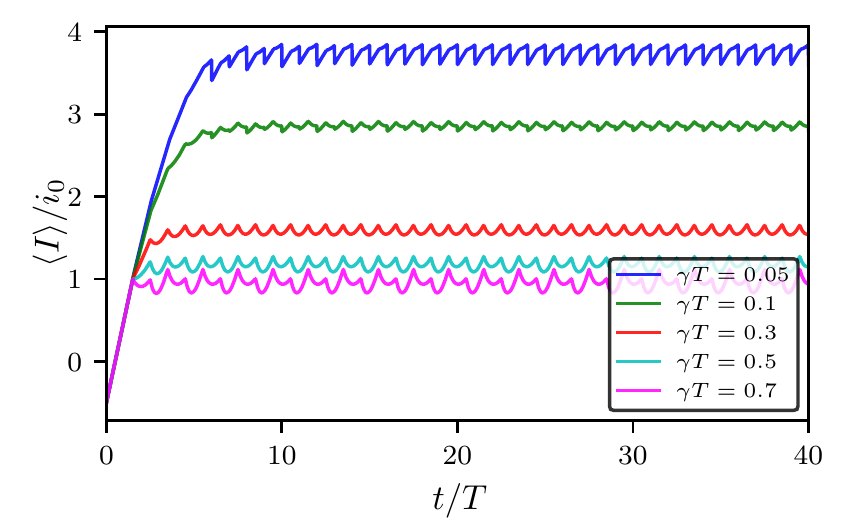}
\caption{(Color online) Mean current as a function of time shows loss of $4\pi$ periodicity at small time-independent dissipation rates in a $2N=10$ level junction.}
\label{fig:meanI_trans}
\end{figure}

\begin{figure}[t]
\includegraphics{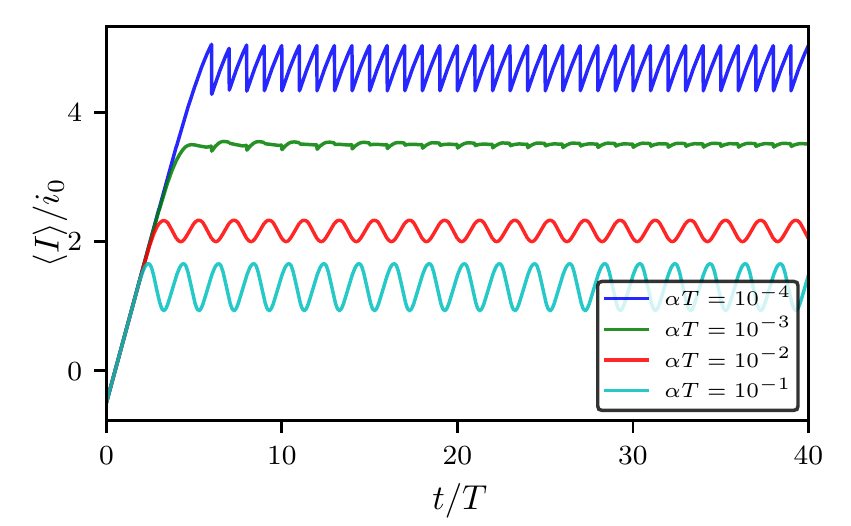}
\caption{(Color online) Mean current as a function of time shows loss of $4\pi$-periodicity in an $2N=10$ level junction at small time-dependent dissipation rates $\alpha$ in units of $(\pi E_T)^3$.}
\label{fig:meanI_td}
\end{figure}

To gain intuition about the power spectrum, we also investigate the mean currents in the transient regime.
The total mean current $\langle I\rangle$ ($I=I_{gs}+I_{ne}$) of the system follows by solving the rate equation and using Eqs.~\eqref{igs} and~\eqref{curr0}.
The $4\pi$ pattern of the Josephson current remains visible in the mean current when the system evolves for time scales below $\tau_{4\pi}$.
We exemplify in a $2N=10$ level junction the loss of $4\pi$ periodicity in the current occurring either for time-independent dissipation (Fig.~\ref{fig:meanI_trans}) or for time-dependent dissipation (Fig.~\ref{fig:meanI_td}).
The current evolution over the first periods is dictated by the time scale required to fill the levels $\tau_\textrm{fill}\approx NT$.
We see that $\tau_{4\pi}$ becomes larger when increasing dissipation rates (more than 40 cycles in Fig.~\ref{fig:meanI_td}), while at lower dissipation rates it quickly decays into a $2\pi$-periodic current.
The amplitude of the mean current may become very small in some cases.
This is due to compensation between the linear increase of the ground-state current within a period and the almost linear decrease of the nonequilibrium current inside a period~[see Figs.~\ref{fig:mean}(a) and (b)].

\begin{figure}[t]
\includegraphics[width=\columnwidth]{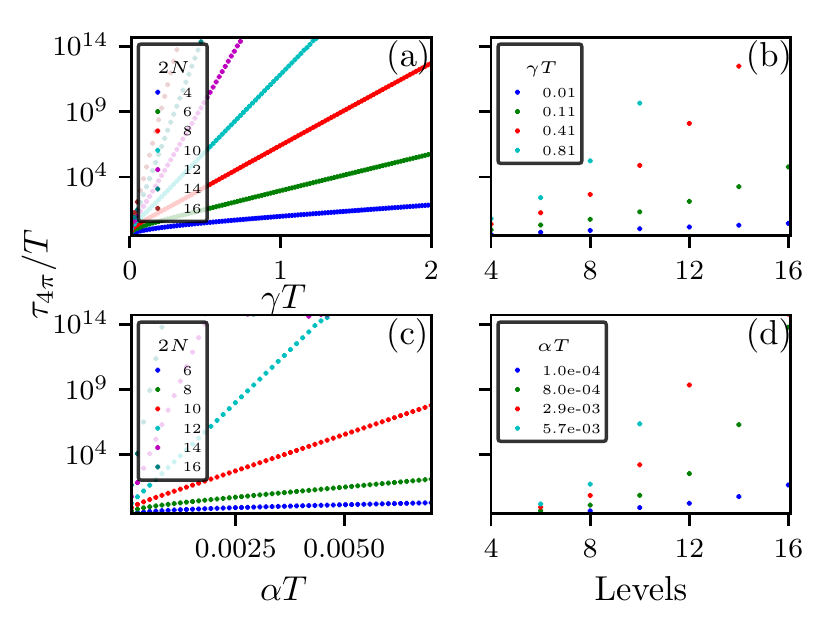}
\caption{(Color online) Lifetime of the $4\pi$-periodic current evaluated from the largest subunitary eigenvalue of the evolution operator.
Panels (a) and (b) stand for energy-independent dissipation and panels (c) and (d) time-dependent relaxation.
The common $y$ axis uses logarithmic scale.
Exponential growth with dissipation strength in (a) and (c).
Panels (b) and (d) show exponential growth with the number of levels for time-independent and, respectively, time-dependent relaxation $\alpha T$ in accordance with the scaling relations in Eq.~\eqref{glarge}, respectively~\eqref{alarge}.
}
\label{fig:lifetime}
\end{figure}

The lifetime of the $4\pi$-periodic current is separately determined from the knowledge of the evolution operator~\eqref{evo} over a period $U(t+T, t)$.
The evolution operator for the rate equation is a Markov matrix with eigenvalues $|\lambda_1|> |\lambda_2|\ge\dots \ge |\lambda_{2N+1}|$.
The unique steady state corresponds to the largest eigenvalue $|\lambda_1|=1$.
The other states are transient and over $n$ periods they decay to the steady state as $|\lambda_{i>1}|^n$.
Therefore, an upper bound estimate of $\tau_{4\pi}$ is given by second largest eigenvalue $\lambda_2$, which controls the decay of the last, most long-lived transient state:
\begin{equation}\label{4pi_eig}
|\lambda_2|^{n}\simeq e^{-nT/\tau_{4\pi}},
\quad \tau_{4\pi}\simeq-\frac{T}{\ln |\lambda_2|}.
\end{equation}
Note that this equation reproduces the short-junction result~\eqref{sjtau4pi}, where $\lambda_2=-r$.
The scaling relations, drawn in the previous subsection, predicting exponential growth of $\tau_{4\pi}$ with dissipation strength and number of levels are now verified directly using Eq.~\eqref{4pi_eig}.
Since the transient state approaches exponentially fast the steady state ($|\lambda_2|\to 1$), the difference between them surpasses quickly the machine precision as either dissipation strength or number of levels increases.
The results are presented in Fig.~\ref{fig:lifetime} for both dissipation models.
At strong dissipation, the results confirm the exponential dependence of $\tau_{4\pi}$ lifetime on dissipation strength and on the number of levels [$\exp(N^3)$ for time-independent dissipation or $\exp(N^6)$ for time-dependent dissipation] from Eqs.~\eqref{glarge} and \eqref{alarge}.

Finally, we compute the power spectrum in both models and show the presence of the fractional peak at strong two-particle dissipation rates.
For time and energy-independent dissipation rates $\gamma$, the expression for the nonequilibrium power spectrum~\eqref{pownene} further simplifies by analytically integrating over the long measurement time~$\tau$ [see Appendix~\eqref{app:pow_ti}].
For time-dependent dissipation rates, the time evolution operator for the state probabilities becomes time dependent.
Consequently, the time integrals in the power spectrum require time-ordered products and the simple expression~\eqref{powsimp} for the power spectrum may no longer be used.
Instead, the power spectrum is determined by numerically propagating the vector of probabilities according to Eqs.~\eqref{steadyp} and~\eqref{pownene}.

\begin{figure}[t]
\includegraphics[width=\columnwidth]{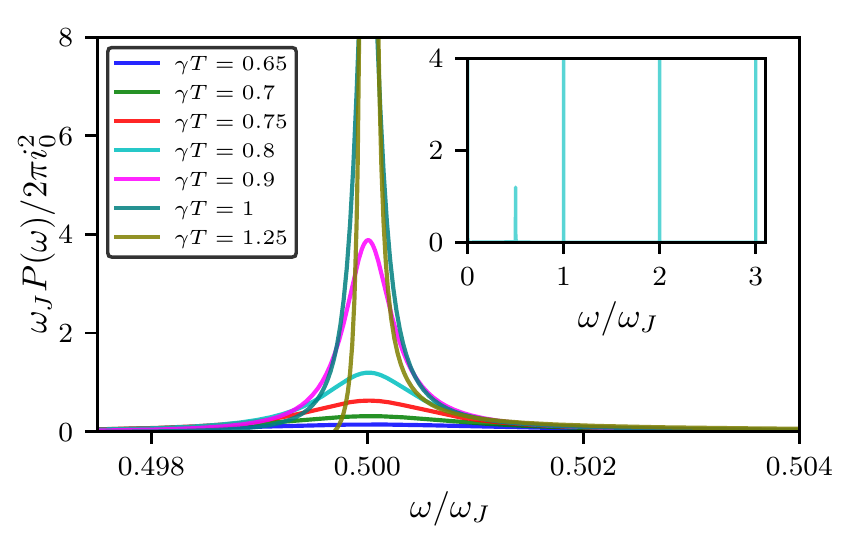}
\caption{(Color online) Fractional peak in the finite-frequency power spectrum for an six-level junction with time and energy-independent relaxation rates $\gamma$.
The inset shows also the usual delta peaks at integer frequencies $\omega/\omega_J$ due to trivial $2\pi$ components of the Josephson current at a given dissipation strength.
The system was evolved over $10^5$ cycles.
}
\label{fig:s}
\end{figure}

\begin{figure}[t]
\includegraphics[width=\columnwidth]{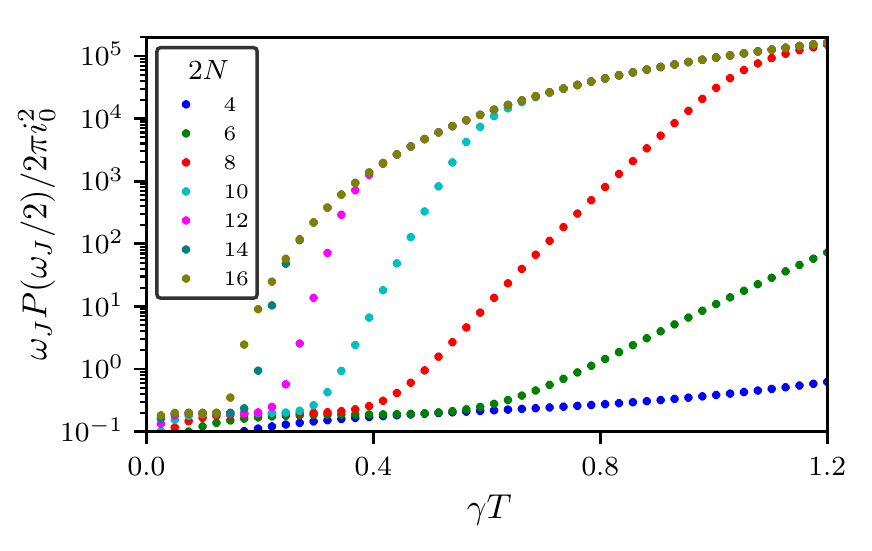}
\caption{(Color online) Height of the fractional peak in the power spectrum for the time and energy-independent dissipation rates $\gamma$ for increasing number $2N$ of levels in the junction.
The peak height is proportional to the lifetime $\tau_{4\pi}$ and shows the predicted exponential growth with dissipation strength.
The peak for any junction deviates at large values from the correct result and saturates due to finite simulation length (here $10^7$ cycles).}
\label{fig:peak}
\end{figure}

\begin{figure}[t]
\includegraphics{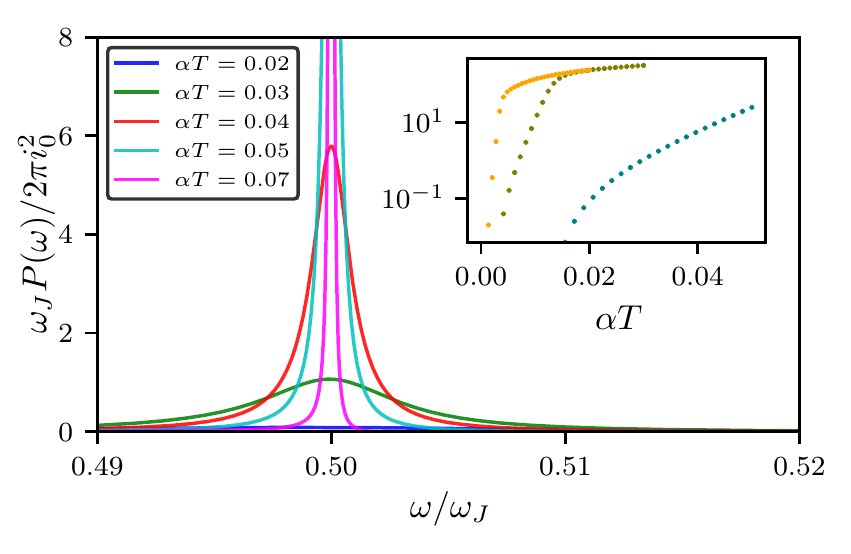}
\caption{(Color online) Fractional peak in the finite frequency power spectrum for a six-level junction in the time and energy-dependent relaxation model at different dissipation strengths $\alpha$ (simulation time span is $10^5$ cycles).
The inset presents the evolution of the fractional peak height (proportional to the lifetime $\tau_{4\pi}$) for 6-, 8-, and 10-level junctions (simulation time span is $10^4$ cycles).
The dissipation strength $\alpha$ is measured in units of $(\pi E_T)^3$.}
\label{fig:powspec_td}
\end{figure}

The power spectrum for time-independent dissipation (Figs.~\ref{fig:s} and~\ref{fig:peak}) and for time- and energy-dependent dissipation (Fig.~\ref{fig:powspec_td}) shows the signature of $4\pi$-periodic Josephson effect in peaks at half-Josephson frequency $\omega/\omega_J=1/2$.
The integer peaks in the power spectrum are also present, as Dirac delta peaks, or diverging with the length of the simulation.
In contrast, the fractional peaks develop at some critical dissipation strength and have a finite width, associated with the lifetime of the $4\pi$-periodic mean current.
For time-independent dissipation we run simulations of $10^5$ Josephson cycles, for increasing number of levels in the junction.
Using Eq.~\eqref{powsimp}, we extract the behavior of the peak corresponding to even longer simulation times (see Fig.~\eqref{fig:peak}).
At relative high dissipation rates or number of levels, the lifetime surpasses the simulation time, leading to an unphysical saturation of the peak height.
The observed exponential dependence on dissipation strength, before saturation, reinforces the previous results from Fig.~\ref{fig:lifetime}.
Moreover the exponential peak develops at some critical dissipation which is indeed lowered with the number of levels increase as suggested by
Eq.~\eqref{gcrit}.
The same conclusions are supported in the case of time- and energy-dependent dissipation rates in Fig.~\ref{fig:powspec_td}.
The fractional peak diverges even faster with dissipation strength and number of levels, as suggested by Eq.~\eqref{alarge}, and the critical dissipation strength is lowered with the number of levels.

\begin{figure}[t]
\includegraphics[width=0.9\columnwidth]{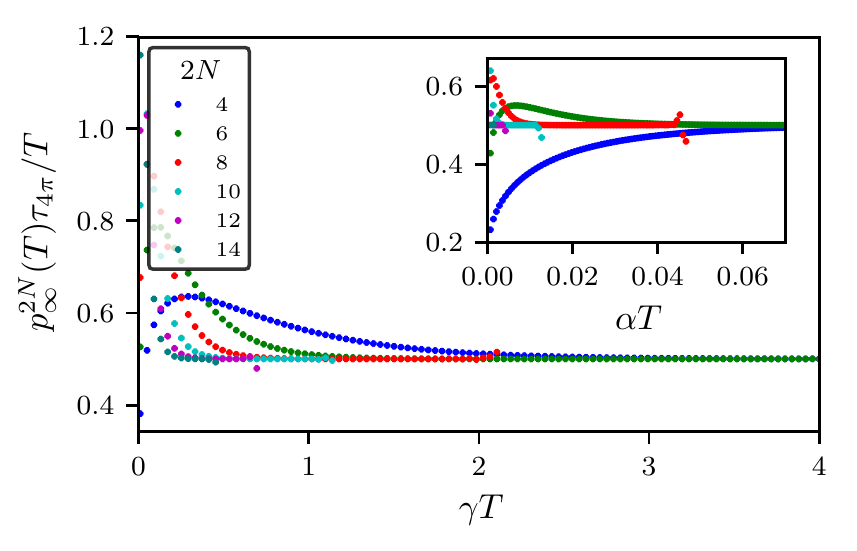}
\caption{(Color online) The fermion parity lifetime $\tau_{4\pi}$ is inverse proportional to the average population of the last level at its entrance into the continuum $p^{2N}_{\infty}(T)$.
Data where $\tau_{4\pi}/T$ exceeds $10^{14}$ are excluded.
The main panel shows the results for the time-independent relaxation model, with the inset, for the time-dependent dissipation model.
The shared legend shows the number of levels in the junctions.}
\label{fig:ratio}
\end{figure}

Our hypothesis, that the lifetime $\tau_{4\pi}$ is inversely proportional with the probability that the particles reach the continuum, is checked once more in the strong dissipation regime.
The particle loss to the continuum due to ejection from the last level is given by the average population of quasiparticles in the steady state, in the highest Andreev level at its entrance into the continuum of states above the gap $p^{2N}_\infty(T)$.
Therefore, the product $\tau_{4\pi}\times p^{2N}_{\infty}$ must tend to a constant, independent on the number of levels.
Figure~\ref{fig:ratio} shows that at strong dissipation, where $\tau_{4\pi}$ is well estimated by the second eigenvalue of the evolution operator over a period, the fermion parity lifetime $\tau_{4\pi}$ is indeed inverse proportional to the population of the last level for both our models for relaxation.

\section{Concluding remarks}
\label{sec:conc}
In this paper we have proved that two-particle relaxation in long QSH junctions generates a long-lived  $4\pi$-periodic Josephson current for two different models of dissipation.
The $4\pi$ periodicity is due to a $4\pi$ periodicity in the fermion parity of the junction.
We have shown how effective two-particle relaxation protects such periodicity as it counteracts the single-particle dissipation events into the continuum of states above the superconducting gap.

The signatures of $4\pi$-periodicity manifest in the junction power spectrum as a peak at half of the Josephson frequency $\omega/\omega_J=1/2$ similar to the one expected in topological junctions supporting Majorana fermions.
These findings offer a possible explanation to the observed signatures of $4\pi$ periodicity in Josephson junctions made of HgTe/CdTe quantum wells~\cite{Deacon2017}.
This hinges, however, on the effectiveness of two-particle dissipation rates in the experiments.
Nevertheless, our proposed mechanism to generate a $4\pi$ periodicity is generic. The two-particle dissipation leads to a dc current in addition to the $4\pi$-periodic current, which has universal (i.e., $N$-independent) voltage dependencies [Eqs.~\ref{eq:idc1} and Eq.~\ref{eq:idc2}].
At low voltages corresponding to the limit of small but universal dc current, we find that the lifetime of the $4\pi$-periodic diverges exponentially, limited only by quasiparticle poisoning and voltage noise for both our relaxation models with the number of levels, two-particle dissipation strength, and inverse Josephson frequency or bias.
This leads to an exponentially higher and sharper fractional Josephson peak in the power spectrum of the current.
Additionally, voltage noise will likely lead to line width broadening, as seen in experiments~\cite{Deacon2017}.
The observation of such correlation between voltage dependence of the dc current as well as the spectral peak height in the long junction limit would be a validation of the fractional Josephson effect.

\begin{acknowledgements}
D.S.~thanks O.~Yudilevich for valuable discussions.
This research was supported by the Netherlands Organization for Scientific Research (NWO/OCW)
as part of the Frontiers of Nanoscience program, and an
ERC Starting Grant.
J.D.S.~acknowledges the funding from Sloan Research Fellowship and NSF-DMR-1555135 (CAREER).
\end{acknowledgements}

\appendix

\section{Circuit damping}
\label{app:circ}
This appendix derives the cubic energy dependence of two-particle relaxation rates due to coupling of the Josephson junction to its electromagnetic environment.

The QSH junction and its environment are modeled following Ref.~\onlinecite{Makhlin2001} which treats dephasing of a superconducting qubit.
The Josephson junction ($S$) plus bath ($B$) are described by the Hamiltonian:
\begin{equation}
H=H_S+H_B+H_{SB},
\end{equation}
where the last term is the Josephson junction coupling to the bath.
The Josephson junction Hamiltonian is expanded near $\varphi_0$:
\begin{eqnarray}
H_S&=&\frac{E_C}{2}n^2+H_J(\varphi)\\
&\approx &
\frac{E_C}{2}n^2+H_J(\varphi_0)+\frac{\hbar}{2e}J(\varphi_0)\delta\varphi+\frac{1}{2}E_J\delta\varphi^2,\notag
\end{eqnarray}
where $J$ is the current in the junction and $E_J$, the Josephson energy.
Without loss of generality, $\varphi_0$ is set to 0, and $\delta\varphi$ is denoted simply by $\varphi$.

A large set of harmonic oscillators indexed by $\alpha$ models the electromagnetic bath:
\begin{equation}
H_B=\frac{1}{2}\sum_{\alpha}\Big( \frac{p_\alpha^2}{m}+m\omega^2_\alpha x_\alpha^2\Big).
\end{equation}

The system-bath Hamiltonian models the coupling between the environment voltage fluctuation $\delta V$ and the charge $n$ on the superconducting leads:
\begin{equation}
H_{SB}= en\delta V =en\sum_{\alpha}\lambda_\alpha x_\alpha,
\end{equation}
with $x_\alpha$ the oscillator displacements.
The coupling constant $\lambda_\alpha$ are effective impedances determined by the bath spectral density,
\begin{equation}
\mathcal J=\frac{\pi}{2m}\sum_\alpha
\frac{\lambda_\alpha^2}{\omega_\alpha}\delta(\omega-\omega_\alpha)=\omega\textrm{Re}[Z_t(\omega)],
\end{equation}
where $Z_t(\omega)=[i\omega C+Z^{-1}(\omega)]^{-1}$, with $Z(\omega)$, the impedance of the environment seen by the junction.

Since the level spacing $\delta$ for the bath's energy levels is very small, the dispersion of the coupling constants is approximated:
\begin{equation}\label{lambdas}
\lambda_\alpha^2\approx 2m\omega^2\textrm{Re}Z_t(\omega)\delta/\pi\hbar.
\end{equation}

The total Hamiltonian, after neglecting the irrelevant shift $H_J(0)$, reads as
\begin{eqnarray}
H&=&\frac{1}{2m}\Big(mE_J\varphi^2+\sum_\alpha p_\alpha^2
\Big)+\frac{m}{2}\Big(\frac{E_C}{m}n^2+\sum_{\alpha}\omega_\alpha^2x_\alpha^2\notag \\
&&+\frac{2en}{m}\sum_\alpha\lambda_\alpha x_\alpha\Big)
+\frac{\hbar}{2e}J\varphi.
\end{eqnarray}

The total Hamiltonian $H$ is more transparently written in vector-matrix notation in a basis of canonically conjugate variables, momentum-like $\eta=(\sqrt{m E_J}\varphi,p_\alpha)^T$ and position-like $\zeta=(\hbar n/\sqrt{m E_J}, x_\alpha)^T$:
\begin{equation}
[\zeta_j,\eta_k]=i\hbar\delta_{jk},
\end{equation}
as
\begin{equation}
H=\frac{1}{2m}\eta^T\eta+\frac{m}{2}\zeta^T M\zeta
+\frac{\hbar}{2e}\frac{\eta_1 J}{\sqrt{mE_J}}.
\end{equation}
The position-mixing matrix $M$ reads as
\begin{equation}
M=\pmat{\Omega^2 & \bm\lambda^T \\
\bm\lambda &\textrm{diag}(\{\omega^2_\alpha\})
},
\end{equation}
where $\hbar\Omega=\sqrt{E_CE_J}$, $\bm\lambda$ is the vector of couplings $e\lambda_\alpha\sqrt{E_J/m}/\hbar$, and $\textrm{diag}(\{\omega_\alpha^2\})$ is a large diagonal matrix of environment oscillator frequencies.

The matrix $M$ is diagonalized $M=UDU^T$, with $D=\textrm{diag}(\Omega^{\prime 2},\{\omega^{\prime2}_\alpha\})$. In new canonically conjugate variables $\eta'=U\eta$ and $\zeta'=U\zeta$, the total Hamiltonian becomes diagonal, except for the ``interaction'' term in $J$:
\begin{equation}\label{diagH}
H=\frac{1}{2m}\eta'^T\eta'+\frac{m}{2}\zeta'^T D\zeta'
+\frac{\hbar}{2e}\frac{(U^T\eta')_1 J}{\sqrt{mE_J}}.
\end{equation}
The last term explicitly reads as
\begin{equation}
\frac{(U^T\eta')_1 J}{\sqrt{mE_J}}
=U_{11}\varphi'J+\sum_\alpha \frac{U_{\alpha 1}}{\sqrt{mE_J}}p_\alpha'J,
\end{equation}
where $U_{1}=(U_{11},U_{\alpha_1 1},U_{\alpha_2 1},\dots)^T$ is the first eigenvector of $M$ with corresponding eigenvalue close to $\Omega^2$.
First-order perturbation theory in small coupling constants $\lambda_\alpha$ determines
\begin{equation}
U_{\alpha 1}
=U_{11}\sqrt{\frac{E_J}{m}}\frac{e\lambda_\alpha/\hbar}
{\Omega^2-\omega_\alpha^2},
\end{equation}
where $U_{11}$ is fixed by requiring that $U_{1}$ is normalized. Remark that to first order $\Omega'=\Omega$ and $\omega'_\alpha=\omega_\alpha$.

Therefore, the interaction term reads as
\begin{equation}\label{int}
H_\textrm{int}=\frac{U_{11}J}{2}
\Big[
\frac{\hbar\varphi'}{e}
+\sum_\alpha\frac{\lambda_\alpha}{m(\Omega^2-\omega_\alpha^2)}p_\alpha'
\Big].
\end{equation}

To compute the two-particle relaxation rates in the junction, we expand the current operator in the basis of Bogoliubov operators:
\begin{equation}
\hat J=\sum_{ij}\Lambda_{ij}c^\dag_i c^\dag_j+\textrm{H.c}+\cdots,
\end{equation}
where only the terms responsible for spin-flip relaxation processes are written explicitly.
We remind again that a term like $c_i c_j$, when, e.g.,~$\sgn(\varepsilon_i)>0$ and $\sgn(\varepsilon_j)<0$, signifies that a right-moving quasiparticle on level $i$ is destroyed and becomes a left-moving quasiparticle in $j$.

The Fermi golden rule determines the two-particle relaxation rate with the Hamiltonian from Eq.~\eqref{int}:
\begin{equation}
\gamma_{ij}\approx
\frac{2\pi}{\delta}\int d\omega_\alpha
\frac{(|\Lambda_{ij}| \lambda_\alpha U_{11}
|\langle \alpha'|p_\alpha'|0\rangle|)^2}
{4m^2(\Omega^2-\omega_\alpha^2)^2}
\delta(\hbar\omega_\alpha-\varepsilon_{ij}),
\end{equation}
where $\varepsilon_{ij}=\varepsilon_i+\varepsilon_j$ is the sum of level $i$ and $j$ energies.
If $\varepsilon_i+\varepsilon_j>0$, then $\gamma_{ij}$ is the rate to annihilate two particles in levels $i$ and $j$ and, if $\varepsilon_i+\varepsilon_j<0$, it is the rate to fill two holes.
Note again that in our convention a particle in a negative-energy level is physically equivalent to an empty positive-energy left-moving state, and a hole at negative energy is physically an excited left-moving quasiparticle.

Substituting Eq.~\eqref{lambdas} for $\lambda_\alpha$ and the matrix element for momentum $p'_\alpha$ yields
\begin{equation}
\gamma_{ij}\approx\frac{|\Lambda_{ij}|^2U_{11}^2
\textrm{Re}[Z_t(\varepsilon_{ij}/\hbar)]
|\varepsilon_{ij}|^3}
{2(\hbar^2\Omega^2-\varepsilon_{ij}^2)^2}.
\end{equation}
The normalization factor $U_{11}$ is on the order 1 in a perturbation theory for small coupling constants.
Further simplifications are available by assuming a resistive inductance $Z(\omega)\approx R$ and  $\Omega\gg|\varepsilon_{ij}|/\hbar$.
Therefore, to first order, the rates read as
\begin{equation}
\gamma_{ij}\approx
\frac{R|\Lambda_{ij}|^2
|\varepsilon_{ij}|^3}
{2\hbar^4\Omega^4(1+R^2C^2\varepsilon_{ij}^2/\hbar^2)}.
\end{equation}

To get the leading behavior for the rates, we approximate the plasma frequency by the superconducting gap $\hbar\Omega\approx\Delta$ and assume that the constant frequency $1/RC$ is longer than the other frequencies in the problem, as for an overdamped junction.

Under these assumptions, we find:
\begin{equation}
\gamma_{ij}(t)\approx \frac{R|\Lambda_{ij}|^2}{2\Delta^4}
|\varepsilon_i(t)+\varepsilon_j(t)|^3,
\end{equation}
where levels $i$ and $j$ are filled in a state $s$ and empty in its descendant state $s'$.

The matrix element $\Lambda_{ij}$ is suppressed by the spin part of matrix elements of the quasiparticle modes and cannot be determined without a microscopic theory.
The matrix element becomes zero in a pristine junction.
Non-idealities allow for a non-zero value of $\Lambda$.

We have thus showed that a good approximation for the two-particle relaxation rates assumes a cubic dependence on the energies of excited quasiparticle levels.

\section{The complete rate equation}
\label{app:full_rate}
This appendix gathers a few results concerning the full rate equation~\eqref{full_rate}, which comprises both spin-conserving and spin-flip relaxation processes.
We show here that the previous results are recovered when spin-conserving processes are much faster than the spin-flip ones, $\chi\gg\gamma$.
While the spin-conserving relaxation processes cannot generate a $4\pi$-periodic current, they help to enhance the visibility of the fractional peak by reducing the probability that quasiparticles escape into the continuum.
To simplify the analysis, we choose here a model with time- and energy-independent relaxation rates $\chi$ (fast spin conserving) and $\gamma$ (slow spin flip).

\begin{figure}[t]
\includegraphics[width=\columnwidth]{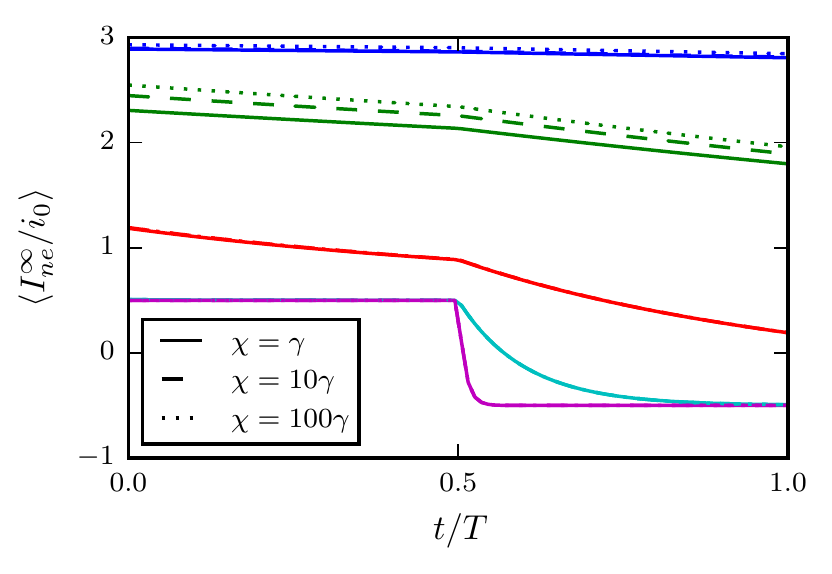}
\caption{(Color online) Steady state for different values of spin-conserving dissipation processes $\chi$ in a six-level junction, when $\gamma T$ varies in the set $\{0.01, 0.1, 1, 10, 100\}$.
At strong spin-flip dissipation, the effect of spin-conserving dissipation processes is negligible.}
\label{fig:steady_slow}
\end{figure}

\begin{figure}[t]
\includegraphics[width=\columnwidth]{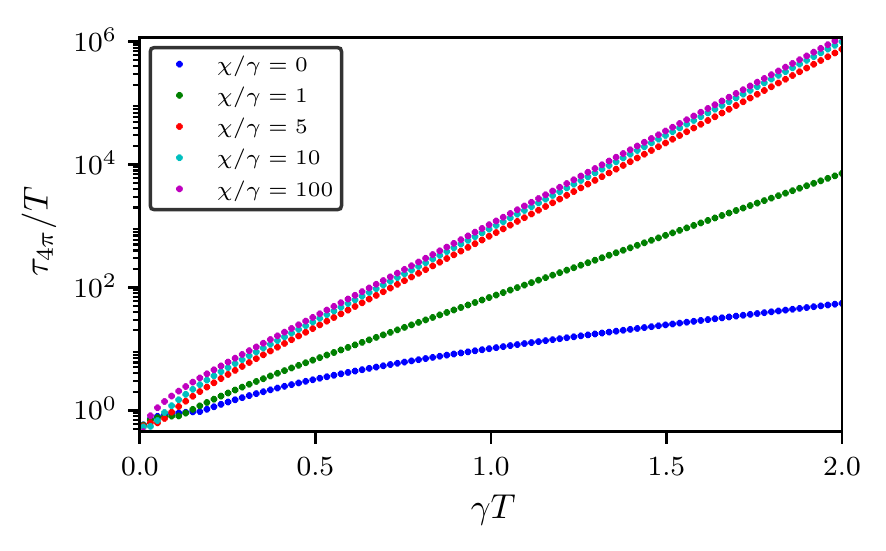}
\caption{(Color online) Lifetime for the $4\pi$-mean current for different ratios of rates $\chi$ and $\gamma$ in a six-level junction.
When spin-conserving processes are fast $\chi/\gamma>1$ the lifetime behavior tends to the result in the fast relaxation approximation.}
\label{fig:t4pi_full}
\end{figure}

\begin{figure}[t]
\includegraphics[width=\columnwidth]{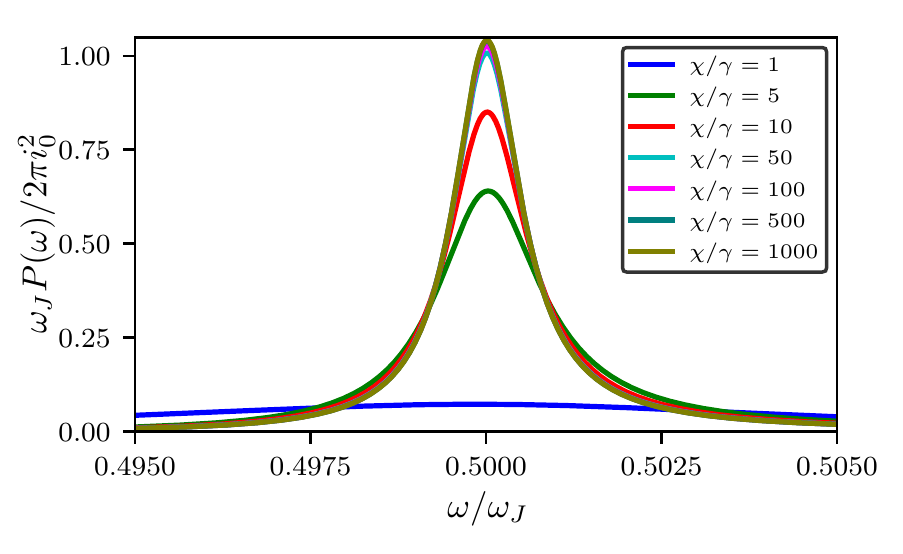}
\caption{(Color online) Typical peak at half the Josephson frequency in the spectrum of a $6$-level junction.
The strength of spin-conserving relaxation processes $\chi$ varies for a given magnitude of spin-flip relaxation strength $\gamma T=0.8$.
Simulation length is $10^5$ cycles.}
\label{fig:peak_full}
\end{figure}

The steady states depend very little on spin-conserving relaxation processes, with differences seen only at small dissipation strengths.
Once the spin-flip particle relaxation becomes relevant $\gamma>\gamma_c$, the effects of spin-conserving relaxation processes become negligible (see Fig.~\eqref{fig:steady_slow}).
In the strong-dissipation limit the mean steady-state current evolves towards the same step distribution as in Eq.~\eqref{mean_strong}.
Since the periodic steady states are unchanged, the same conclusions also hold for dc currents as in Sec.~\ref{sec:Nlev}.

The lifetime $\tau_{4\pi}$ corresponding to the decay of most long-lived transient state Eq.~\eqref{4pi_eig} is presented in Fig.~\ref{fig:t4pi_full} for a 6-level junction (with similar behavior observed for 8- and 10-level junctions).
When $\chi$ processes are faster than $\gamma$ (which is physically the case) $\chi/\gamma>1$, the dependence of the lifetime on dissipation strength $\gamma$ becomes exponential as in the fast relaxation approximation and approaches the known results displayed in Fig.~\ref{fig:lifetime}(a).

Finally, we compute the power spectrum at some known dissipation $\gamma$ in a six-level junction, but now including the effects of spin-flip dissipation processes.
At low spin-flip rates $\chi$, the fractional peak in the spectrum is small and broad.
As $\chi$ becomes larger than $\gamma$, spin-conserving processes help preventing quasiparticles escape into continuum and spin-flip processes become effective in generating the $4\pi$ currents.
The fractional peak shown in Fig.~\ref{fig:peak_full} recovers the results from Fig.~\ref{fig:peak} when spin-conserving processes become much faster on the scale of spin-flip processes $\chi\gg\gamma$.

\section{Power spectrum for time-independent dissipation}
\label{app:pow_ti}
The general expression for the nonequilibrium power spectrum~\eqref{pownene} further simplifies when considering time and energy-independent dissipation rates, by first assuming that the long-time interval over which measurement is carried contains a large integer number of $M$ periods, $\tau=MT$.
The rate matrix is different in each half of a period due to the possibility of having a positive hole involved in two-hole annihilation processes in the first period, and a negative-energy particle involved in two-particle annihilation processes [Eqs.~\eqref{full_rate} or~\eqref{approx_rate}].
Nevertheless, the dissipation matrix is constant in time in each half of a period:
\begin{equation}
\bm \Gamma(t)=
\begin{cases}
\Gamma_1 & \textrm{frac}(t/T) < 1/2\\
\Gamma_2 & \textrm{frac}(t/T) \ge 1/2
\end{cases},
\end{equation}
with $\Gamma_i$ not commuting with each other.
The evolution operator within any period is denoted by
$U(T)=e^{\Gamma_2\frac T 2}e^{\Gamma_1\frac T 2}$.
Integrating over the measurement time $\tau$ yields the nonequilibrium power spectrum:
\begin{widetext}
\begin{eqnarray}\label{powsimp}
P_{ne\text{-}ne}(\omega)
&=& \frac{2}{T}\text{Re}\bigg\{
\int_0^{\frac T 2} dt\, {\bm I}^T_{ne}\cdot
\bigg[
\frac{e^{\Gamma_1\frac T 2+i\omega(\frac T 2-t)
}-e^{\Gamma_1 t}}{i\omega+\Gamma_1}
+\frac{e^{i\omega(T-t)}-e^{-\Gamma_2\frac T 2
+i\omega(\frac T 2-t)}}{i\omega+\Gamma_2}U
+\frac{e^{\Gamma_1 t+i\omega T} - e^{i\omega(T-t)}}{i\omega+\Gamma_1}
WU
\bigg]\notag\\
&&\cdot \sum_{k=0}^{M-1}{(WUe^{i\omega T})^k}
\cdot e^{-\Gamma_1 t}
\cdot[\bm I_{ne}\circ \bm p_\infty(t)]\\
&&+\int_{\frac T 2}^T dt\, \bm I^T_{ne}\cdot
\bigg[
\frac{e^{i\omega(T-t)}-e^{\Gamma_2(t-T)}}
{i\omega+\Gamma_2}
+\frac{e^{\Gamma_1\frac{T}{2}+i\omega(\frac{3T}{2}-t)} - e^{i\omega(T-t)}}{i\omega+\Gamma_1}W
+\frac{e^{\Gamma_2(t-T)+i\omega T}
-e^{-\Gamma_2\frac T 2 +i\omega(\frac{3T}{2}-t)}}
{i\omega+\Gamma_2} UW
\bigg]\notag\\
&&\cdot \sum_{k=0}^{M-1}{(UWe^{i\omega T})^k}
\cdot e^{\Gamma_2(T-t)}
\cdot[{\bm I}_{ne}\circ \bm p_\infty(t)]
\bigg\}\notag
\end{eqnarray}
\end{widetext}
with $\circ$ denoting the Hadamard (element-wise) product and the nonequilibrium current vector ${\bm I}_{ne}=i_0(\bm n-N\bm 1)$, with $\bm n$ the state occupation vector and $\bm 1^T=(1,1,\dots,1)$.
The geometric sum diverges at integer frequencies $\omega=n\omega_J$, $n\in\mathbb Z$, because the evolution operators $WU$ and $UW$ have one as an eigenvalue.
At large dissipation or for large number of levels, $\tau_{4\pi}$ diverges, which leads to additional divergences at fractions of the Josephson frequencies $\omega=(2n+1)\omega_J/2$.
At any other frequencies it is safe to perform the summation over $k$ and take the limit $M\to\infty$ to obtain $(1-UWe^{i\omega T})^{-1}$ or $(1-WUe^{i\omega T})^{-1}$.

Compared to Eq.~\eqref{pownene}, the expression~\eqref{powsimp} allows to evolve the system over longer times, thus improving the power spectrum resolution in frequency.
\bibliographystyle{apsrev4-1}
\bibliography{bibl}
\end{document}